\documentclass[reprint,superscriptaddress,showpacs,preprintnumbers,nofootinbib,amsmath,amssymb,floatfix,aps,prd,twocolumn]{revtex4-1}
\usepackage{graphicx}
\usepackage{subfigure}%
\usepackage{dcolumn}
\usepackage{isotope}
\usepackage{multirow}
\usepackage{color}
\usepackage{hyperref}

\newcommand{\ve}[1]{\ensuremath{\mathbf{#1}}}
\newcommand{\n}[1]{\ensuremath{|\mathbf{#1}|}}

\newcommand{\Ekp}{\ensuremath{E_{\ell}}}
\newcommand{\Ep}{\ensuremath{E_{\mathbf p}}}
\newcommand{\Eppi}{\ensuremath{E_{{\mathbf p}_i'}}}
\newcommand{\Epp}{\ensuremath{E_{{\mathbf p'}}}}
\newcommand{\GENIE}{\textsc{genie}}
\newcommand{\GLoBES}{\textsc{gl}{\small o}\textsc{bes}}

\newcommand{\ie}{i.e., }

\begin{document}

\preprint{FERMILAB-PUB-15-301-T}
\preprint{RM3-TH/15-10}

\title{Comparison of the calorimetric and kinematic methods\\ of neutrino energy reconstruction in disappearance experiments}

\author{A. M. Ankowski}
\email{ankowski@vt.edu}
\affiliation{Center for Neutrino Physics, Virginia Tech, Blacksburg, Virginia 24061, USA}
\author{O. Benhar}
\affiliation{INFN and Department of Physics,``Sapienza'' Universit\`a di Roma, I-00185 Roma, Italy}
\affiliation{Center for Neutrino Physics, Virginia Tech, Blacksburg, Virginia 24061, USA}
\author{P. Coloma}
\affiliation{Fermi National Accelerator Laboratory, Batavia, Illinois 60510, USA}
\author{P. Huber}
\affiliation{Center for Neutrino Physics, Virginia Tech, Blacksburg, Virginia 24061, USA}
\author{C.-M.~Jen}
\affiliation{Center for Neutrino Physics, Virginia Tech, Blacksburg, Virginia 24061, USA}
\author{C. Mariani}
\affiliation{Center for Neutrino Physics, Virginia Tech, Blacksburg, Virginia 24061, USA}
\author{D. Meloni}
\affiliation{INFN and Dipartimento di Matematica e Fisica, Universit\`a di Roma Tre, Via della Vasca Navale 84, 00146 Rome, Italy}
\author{E. Vagnoni}
\affiliation{INFN and Dipartimento di Matematica e Fisica, Universit\`a di Roma Tre, Via della Vasca Navale 84, 00146 Rome, Italy}

\begin{abstract}
To be able to achieve their physics goals, future neutrino-oscillation
experiments will need to reconstruct the neutrino energy with very
high accuracy. In this work, we analyze how the energy reconstruction
may be affected by realistic detection capabilities, such as energy
resolutions, efficiencies, and thresholds. This allows us to estimate
how well the detector performance needs to be determined \emph{a
  priori} in order to avoid a sizable bias in the measurement of the
relevant oscillation parameters. We compare the kinematic and
calorimetric methods of energy reconstruction in the context of two
$\nu_\mu \rightarrow \nu_\mu$ disappearance experiments operating in
different energy regimes.  For the calorimetric reconstruction method,
we find that the detector performance has to be estimated with an
$\mathcal{O}(10\%)$ accuracy to avoid a significant bias in the
extracted oscillation parameters. On the other hand, in the case of
kinematic energy reconstruction, we observe that the results exhibit
less sensitivity to an overestimation of the detector capabilities.
\end{abstract}


\maketitle

\section{Introduction}

Long-baseline experimental searches of neutrino oscillations largely
rely on the capability of pinning down the energy dependence of the
oscillation probability, which is a nontrivial function of the {\em
  true} neutrino energy, $E_\nu$. As a consequence, the procedure
employed to reconstruct the unknown incoming-neutrino energy from the
measured kinematics of the interaction products is a central
element of the oscillation analysis.

Experiments using neutrino beams peaked at $E_\nu \sim  600$--800 MeV, such as T2K~\cite{ref:T2K} and MiniBooNE~\cite{ref:MiniBooNE},  determine
the energy distribution of charged-current (CC) events from the
kinematics of the outgoing charged lepton---\ie its
kinetic energy and emission angle---measured by large Cherenkov
detectors filled with water or mineral oil. This technique is mostly
applied to quasielastic (QE) events---identified by the absence of
pions in the final state---that provide the dominant contribution to
the total cross section at these energies. However,
it necessarily involves hypotheses on the reaction mechanism.

The kinematic method of energy reconstruction is based on the assumptions that the beam
particle interacts with a single nucleon at rest, bound with constant
energy, and that no other nucleons are knocked out from the nucleus.
It has been long known, however, that processes involving two-nucleon currents, final-state
interactions, and nucleon-nucleon correlations give rise to the
appearance of more complex final states, featuring more than one
nucleon excited to the continuum. Reconstruction of the neutrino
energy of such events, dubbed QE-like in the mesonless case~\cite{ref:Martini_QElike}, in
general requires more complex methods, involving realistic models
of nuclear dynamics.

At energies well above 1 GeV, the contribution of inelastic
processes---resonant pion production and deep-inelastic scattering
(DIS)---becomes larger and eventually dominant. As a consequence, determination of the neutrino energy in this kinematic region requires reconstruction of events
with many hadrons in the final state.

As alternative to Cherenkov detectors in this regime, calorimeters
measuring the visible energy associated with each event---\ie the
energy deposited by the final-state particles---have been proposed as
effective devices, allowing for an accurate neutrino-energy
reconstruction.
Calorimeters are presently being used in the MINOS~\cite{ref:MINOS} and NO$\nu$A~\cite{ref:NOvA} experiments.
In their detectors, the total energy deposited by all reaction products is measured
without a prior reconstruction of each final-state particle's track,
momentum, or energy. The energy response for potentially complex
final states is calibrated using test-beam exposures.

On the other hand, future appearance experiments like DUNE~\cite{ref:LBNE} are expected to employ
detectors capable of fine-grained tracking of a large number of interaction products.
In their case, the tracking capability is the key to being able to select electron-neutrino events and distinguish them
from backgrounds, even for non-QE events.

In this article, we compare the
kinematic and calorimetric reconstruction methods in the oscillation
analysis of a $\nu_\mu$ disappearance experiment. We also aim to explore
the capabilities and limitations of a calorimetric analysis based on
individually identified particle tracks and to determine what level of
understanding of detector response and underlying events is required to
meet certain physics goals.

For a muon-neutrino
event, it is clear that the long muon track in itself is a clear signature, and no
tracking beyond the leading muon is required for backgrounds removal.
However, we study the $\nu_\mu$ disappearance channel because of its
simplicity in terms of oscillation physics, and as a sandbox to develop
suitable analysis tools.

The calorimetric technique, as defined above, obviously rests on the
ability of fully reconstructing the final state, which largely depends
on the detector design and performance. Nuclear effects also play a
role, as they may lead to a sizeable amount of {\em missing} energy,
hindering the reconstruction of $E_\nu$.  For example, if a pion
produced at the elementary interaction vertex is absorbed in the
spectator system, in general, its energy is not deposited in the
calorimeter.

We consider an idealized setting, in which the near and far
detectors are functionally identical, and a simple extrapolation
between them can be performed~\cite{ref:Gallagher}. To minimize the
uncertainty arising from nuclear interactions, the target nucleus
selected for both detectors is carbon, the cross section of which has
been extensively measured in a number of different
channels~\cite{ref:NOMAD_inclusive,ref:NOMAD,ref:SciBooNE_inclusive,ref:MINERvA_anu,ref:MiniB_NC,
  ref:MiniB_CCQE_nu,ref:MiniB_piC,ref:MiniB_pi0,ref:T2K_CC_numu_xsec_C,ref:MINERvA_nu,
  ref:MiniB_CCQE_anu,ref:T2K_CC_numu_xsec_CH,ref:T2K_CCQE_numu_xsec_C,ref:T2K_CCQE_numu_C,
  ref:T2K_CC_nue,ref:MiniB_NCQE_anu,ref:MINERvA_nu_p,ref:MINERvA_CC_ratios,
  ref:MINERvA_coherent_piC,ref:MINERvA_piC,ref:MINERvA_pi0}.

The analyzed events are generated using the simulation code \GENIE{}~\cite{ref:GENIE} supplemented with the $\nu T$ package of
additional modules~\cite{ref:vT}, allowing us to describe the carbon ground state using the realistic spectral function~\cite{ref:Omar_LDA}.

To cover different experimental configurations, we consider two options: a low-energy (LE) setup, with a narrow-band off-axis beam, and a high-energy (HE) setup,
featuring a  broadband on-axis  beam. In the LE option, the neutrino flux is peaked around 600 MeV, and the distance to the far detector is set to $L =295$ km, while in the HE
option, the flux is peaked at $\sim$1--2 GeV and $L = 1000$ km.

Event reconstruction is carried out assuming:
\begin{enumerate}
\item[(i)] a {\em perfect}
scenario, in which all produced particles are detected and their true
energies are measured,
\item[(ii)] a {\em realistic} one, in which
detection efficiencies and thresholds are taken into account and the
finite detector resolution leads to a smearing of the measured
energies.
\end{enumerate}
 In both cases, neutrons are assumed to escape
detection altogether.

A key element of our analysis is the migration matrices, the
elements of which, ${\mathcal M}_{ij}$, correspond to the probability that an
event with a true neutrino energy in the $j$th bin ends up being
reconstructed in an energy bin~$i$.

The event distribution representing the data is in each case generated using the realistic scenario.
Fitted rates, on the other hand, are obtained using a linear combination of migration matrices
corresponding to the perfect and realistic scenarios. This procedure is
very effective from the computational point of view, and provides a
measure of the impact of detector performance on the oscillation-parameters fit.

This article is structured as follows. In
Sec.~\ref{sec:energyReconstruction} we derive different methods of
neutrino-energy reconstruction, placing a special emphasis on the
approximations involved. The neutrino cross section model employed for
event generation and the treatment of detector effects are outlined in
Secs.~\ref{sec:GENIE} and~\ref{sec:scenarios}, respectively.
In Sec.~\ref{sec:migrationMatrices}, we discuss the
calculated migration matrices.  Our oscillation analysis and its
results are presented in Sec.~\ref{sec:oscillationAnalysis}.  Finally, in
Sec.~\ref{sec:summary}, we summarize our findings.

\section{Energy reconstruction methods}
\label{sec:energyReconstruction}

Consider CC neutrino scattering off a nuclear target, resulting in the
knockout of $n$ nucleons and production of $m$ mesons.  The energy
and momentum conservation can be cast in the form
\begin{eqnarray}
E_\nu+M_A&=&\Ekp+E_{A-n}+\sum_{i}\Eppi+\sum_{j}E_{{\mathbf h}_j'},\label{eq:energyConservation}\\
\ve k_\nu&=&\ve k_\ell-\ve{p}+\sum_{i}\ve p'_i+\sum_{j}\ve h'_j\label{eq:momConservation},
\end{eqnarray}
respectively, where $E_\nu$ and $\ve k_\nu$ ($\Ekp$ and $\ve k_\ell$)
are the neutrino's (charged lepton's) energy and momentum, $\Eppi$ and
$\ve p'_i$ denote the energy and momentum of the $i$th knocked-out
nucleon ($1\leq i\leq n$), and $E_{{\mathbf h}_j'}$ and $\ve h'_j$
stand for the energy and momentum of the $j$th produced meson
($1\leq j\leq m$).  The energy of the residual $(A-n)$-nucleon system,
$E_{A-n}$, can be conveniently expressed as
\begin{equation}\label{eq:spectatorSysEnergy}
E_{A-n}=M_A-nM+E+T_{A-n},
\end{equation}
in terms of the nucleon (target-nucleus) mass $M$ ($M_A$), the recoil
energy $T_{A-n}$, and the excitation energy $E$.  In
Eq.~\eqref{eq:momConservation}, the recoil momentum of the system is
denoted as $-\ve p$, to allow the interpretation of $\ve p$ as the
vector sum of the initial momenta of the knocked-out nucleons,
assuming that it is not altered by final-state interactions with the
residual system.

Substitution of Eq.~\eqref{eq:spectatorSysEnergy} into
Eq.~\eqref{eq:energyConservation} leads to the neutrino energy in the
form
\begin{equation}\label{eq:neutrinoEnergy}
E_\nu=\Ekp+E+T_{A-n}+\sum_{i}(\Eppi-M)+\sum_{j}E_{{\mathbf h}_j'}.
\end{equation}
Note that while for mesons the total energies enter the sum, for nucleons only the kinetic energies contribute. This difference is a consequence of the fact that mesons are produced in the interaction process, whereas nucleons are only knocked out from the target nucleus.

Assuming that multinucleon effects do not introduce strong energy dependence to the cross sections, the factor $\epsilon_n=E+T_{A-n}$ can be treated as a constant at neutrino energies above several hundred MeV. Then, reconstruction of the neutrino energy reduces to determining the energies of the particles in the final state~\cite{ref:MINOS_PRL},
\begin{equation}\label{eq:calEnergy}
E^\textrm{cal}_\nu=\epsilon_n+\Ekp+\sum_{i}(\Eppi-M)+\sum_{j}E_{{\mathbf h}_j'}.
\end{equation}

This, so-called, calorimetric method can, in principle, be applied to any type of CC interaction. However, one needs to keep in mind that an accurate reconstruction of hadrons poses a formidable experimental challenge. In particular, neutrons typically escape detection, and any undetected meson results in energy underestimation by at least the value of the pion mass, $\sim$135 MeV.

When the invariant hadronic mass squared, defined as
\[
W^2=\Big(\sum_{i}\Eppi+\sum_{j}E_{{\mathbf h}_j'}\Big)^2-\Big(\sum_{i}\ve p'_i+\sum_{j}\ve h'_j\Big)^2,
\]
is known, Eqs.~\eqref{eq:energyConservation} and \eqref{eq:momConservation} can be solved for the neutrino energy, yielding the alternative expression~\cite{ref:Omar&Davide}
\begin{equation}\label{eq:generalRecEnergy}
E_\nu=\frac{2\Ep \Ekp-2\ve p\cdot \ve k_\ell+W^2-(\Ep^2-\ve p^2)-m_\ell^2}{2(\Ep-\n{p}\cos\theta_{N}-\Ekp+\n{k_\ell}\cos\theta)},
\end{equation}
where $\Ep=M_A-E_{A-n}$, $m_\ell$ is the charged lepton's mass, $\n p\cos\theta_N=\ve p\cdot\ve k_\nu/E_\nu$, and $\n{ k_\ell}\cos\theta=\ve k_\ell\cdot\ve k_\nu/E_\nu$.

On the other hand, in the case of a~single-nucleon knockout associated with the production of $m$ mesons, the requirement that the nucleon is on the mass shell, $\Epp^2=M^2+\ve p'^2$, can be used to obtain the neutrino energy as
\begin{equation}\label{eq:piNRecEnergy}
E_\nu={\frac{M^2+(\ve p-\ve k_\ell-\sum_{j}\ve {h}'_j)^2-(\Ep-\Ekp-\sum_{j}E_{{\mathbf h}_j'})^2}{2\big[\Ep-\n{p}\cos\theta_{N}-\Ekp+\n{k_\ell}\cos\theta-\sum_{j}{\mathcal H}_j\big]}},
\end{equation}
with ${\mathcal H}_j=E_{{\mathbf h}_j'}-\ve h'_j\cdot\ve k_\nu/E_\nu$.

In practice, an application of the above formulas requires (i) neglecting the unmeasured recoil momentum $\n p$ and (ii) approximating the energy of the residual nuclear system by a constant, which amounts to setting $\Ep=nM-\epsilon_n$ in Eq.~\eqref{eq:generalRecEnergy} and $\Ep=M-\epsilon$ in Eq.~\eqref{eq:piNRecEnergy}. These simplifications lead to the expressions \begin{equation}\label{eq:kinRecEnergy_W}
E_\nu^\textrm{kin}=\frac{2(nM-\epsilon_n) \Ekp+W^2-(nM-\epsilon_n)^2-m_\ell^2}{2(nM-\epsilon_n-\Ekp+\n{k_\ell}\cos\theta)},
\end{equation}
and
\begin{equation}\label{eq:kinRecEnergy_multitrack}
E_\nu^\textrm{mt}=\frac{M^2+(\ve k_\ell+\sum_{j}\ve {h'_j})^2-(M-\epsilon-\Ekp-\sum_{j}E_{{\mathbf h}_j'})^2}{2(M-\epsilon-\Ekp+\n{k_\ell}\cos\theta-\sum_{j}{\mathcal H}_j)}.
\end{equation}

Owing to difficulties with an accurate determination of the invariant hadronic mass, the use of Eq.~\eqref{eq:kinRecEnergy_W} is usually restricted to the process of a single-nucleon knockout with no pions produced, in which $W^2$ is known and equals $M^2$~\cite{ref:K2K_PRL,ref:MiniB_kappa,ref:NOMAD,ref:SciBooNE_inclusive,ref:MINERvA_anu}. On the other hand, Eq.~\eqref{eq:kinRecEnergy_multitrack} has been applied in the energy reconstruction for single-pion events by the MiniBooNE Collaboration~\cite{ref:MiniB_piC}, setting the single-nucleon separation energy $\epsilon$ to zero.

The kinematic energy reconstruction, by means of Eq.~\eqref{eq:kinRecEnergy_W} or \eqref{eq:kinRecEnergy_multitrack}, does not require the knocked-out nucleon's momentum to be measured. However, as this method assumes specific final states, its accuracy is spoiled by any undetected hadron.
For example, when a produced pion is absorbed or undetected, the energy reconstructed from Eq.~\eqref{eq:kinRecEnergy_W} under the QE hypothesis is typically lower than the true one by $\sim$300--350 MeV; see Figs.~6 and 7 of Ref.~\cite{ref:Leitner}. While the process of multinucleon knockout affects the energy reconstruction in a~similar way---\ie it redistributes the strength of the reconstructed flux from the peak mainly to the low-energy tail~\cite{ref:Martini_Erec1,ref:Martini_Erec2}---this effect seems to be less relevant at higher beam energies~\cite{ref:Lalakulich_Erec}.

For completeness, we mention that neutrino energy can be also found exploiting momentum conservation only. Equation~\eqref{eq:momConservation} multiplied by a factor $\ve k_\nu/E_\nu$,
\begin{equation}
E_\nu=\n{ k_\ell}\cos\theta-\n p\cos\theta_N+\sum_{i}|{\bf p}'_i|\cos\theta_i+\sum_{j}|{\bf h}'_j|\cos\theta_j,
\end{equation}
where $|{\bf p}'_i|\cos\theta_i=\ve p'_i\cdot\ve k_\nu/E_\nu$ and $|{\bf h}'_j|\cos\theta_j=\ve h'_j\cdot\ve k_\nu/E_\nu$, shows that $E_\nu$ can be determined from the projections of the final momenta on the beam direction. Neglecting the contribution of the recoil momentum, one obtains the expression
\begin{equation}
E^\textrm{mom}_\nu=\n{ k_\ell}\cos\theta+\sum_{i}|{\bf p}'_i|\cos\theta_i+\sum_{j}|{\bf h}'_j|\cos\theta_j,
\end{equation}
a special case of which has been employed in an analysis of single-nucleon knockout events by the NOMAD Collaboration~\cite{ref:NOMAD}.

Performing the kinematic energy reconstruction in this article, we employ Eq.~\eqref{eq:kinRecEnergy_W} assuming single-nucleon knockout, regardless of the actual number of nucleons in the final state. We set $W^2$ to $M^2$ for mesonless events and to $M_\Delta^2$, $M_\Delta= 1.232$ GeV being the $\Delta$ resonance mass, when at least one meson is \emph{observed}. The single-nucleon separation energy is fixed to 34 MeV. The same value of $\epsilon$ is added in the calorimetric energy reconstruction~\eqref{eq:calEnergy} for every nucleon detected.

\section{Event generation}
\label{sec:GENIE}

\begin{figure*}
\centering
    \includegraphics[width=0.80\textwidth]{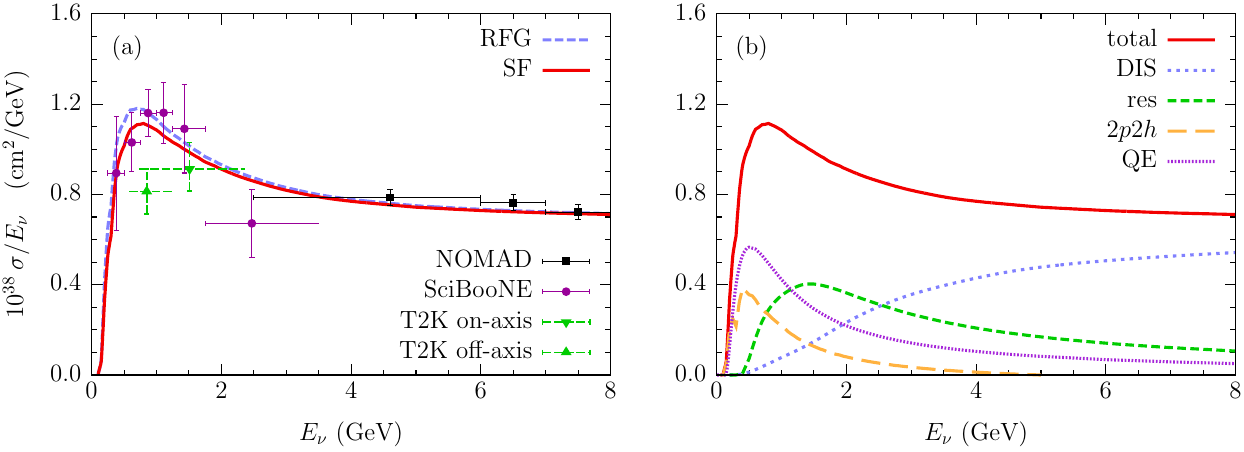}
    \subfigure{\label{fig:xsec_inclusive}}
    \subfigure{\label{fig:xsec_breakup}}
\caption{\label{fig:xsec}(color online). Panel (a): Per-nucleon CC inclusive $\nu_\mu$ cross section divided by neutrino energy, obtained using \GENIE{} $2.8.0+\nu$T~\cite{ref:GENIE,ref:vT} with the relativistic Fermi gas model (dashed line) and the spectral function approach (solid line) as a nuclear model in QE interaction. The results for carbon are compared to the experimental data for carbon extracted from NOMAD~\cite{ref:NOMAD_inclusive} and those for hydrocarbon (CH) reported from SciBooNE~\cite{ref:SciBooNE_inclusive} and T2K~\cite{ref:T2K_CC_numu_xsec_C,ref:T2K_CC_numu_xsec_CH}. Panel (b): Breakup of the contributions to the inclusive cross section. The labels DIS, res, $2p2h$, and QE refer to deep-inelastic scattering, resonant pion production, two-nucleon knockout, and quasielastic scattering, respectively.}
\end{figure*}

Our analysis is based on the description of nuclear structure and interaction dynamics in
the \GENIE{} Monte Carlo generator~\cite{ref:GENIE}, version 2.8.0, supplemented with the $\nu T$ package of additional modules~\cite{ref:vT}.

\GENIE{} is a modern and versatile platform for neutrino event simulation. It has been developed putting special emphasis on scattering in the energy region of few GeV---important for ongoing and future oscillation studies---where various mechanism of interaction are relevant. A number of neutrino experiments employs this generator in data analysis~\cite{ref:Dytman_NuFact10}.

Resonant pion production in \GENIE{}, considered for $W\leq1.7$ GeV, is accounted for using the model of Rein and Sehgal~\cite{ref:Rein&Sehgal}. Compared to 18 in the original model, 16 resonances of unambiguous existence are implemented using up-to-date parameters but neglecting the interference between them. The effect of the charged lepton's mass is taken into account only in the calculations of the phase-space boundaries.

The contribution of nonresonant processes, classified in \GENIE{} as DIS, is calculated following the method of Bodek and Yang~\cite{ref:Bodek&Yang,ref:Bodek&Yang_updated}. This effective approach extends the range of applicability of the parton model to low neutrino energy by modifying the (leading-order) parton-distribution functions in the low-$Q^2$ region, $Q^2$ being the four-momentum transfer squared. Higher-order and target-mass corrections are accounted for by replacing Bjorken $x$ with a new scaling variable~\cite{ref:Bodek&Yang,ref:Bodek&Yang_updated}. While DIS in \GENIE{} is the only mechanism of interaction at $W>1.7$ GeV, it also produces one- and two-pion events in the resonance region.

Hadronization in \GENIE{} is performed using the Andreopoulos--Gallagher--Kehayias--Yang approach~\cite{ref:AGKY}, which combines Koba--Nielsen--Olesen (KNO) scaling~\cite{ref:KNO} at low values of the hadronic invariant mass with the \textsc{pythia/jetset} calculations~\cite{ref:PYTHIA} at high $W$, ensuring a smooth transition between the two regimes.

Two-nucleon knockout ($2p2h$) events are simulated in \GENIE{} using the procedure of Dytman~\cite{ref:GENIE_2p}, obtained modifying and extending the one of Ref.~\cite{ref:Lightbody&OConnell}, derived for electron scattering. The invariant mass of two-nucleon events is assumed to have a Gaussian distribution centered at $W=(M+M_\Delta)/2$. The charged lepton's kinematics is distributed according to the magnetic contribution to the cross section for scattering on a free nucleon~\cite{ref:LlewellynSmith}, identical for neutrinos and antineutrinos. The strength, adjusted to fit the cross sections reported by MiniBooNE~\cite{ref:MiniB_CCQE_nu,ref:MiniB_CCQE_anu}, is implemented to linearly decrease to zero for the neutrino energy between 1 and 5 GeV, to avoid inconsistency with the results from NOMAD~\cite{ref:NOMAD}.
As the experimental data constrain the two-nucleon contribution for the carbon target only, for other nuclei its strength is assumed to exhibit linear dependence on the mass number $A$.

In QE scattering, the default nuclear model in \GENIE{} is the relativistic Fermi gas (RFG) model of Bodek and Ritchie~\cite{ref:Bodek&Ritchie}, in which a high-momentum tail---inspired by the effects of nucleon-nucleon correlations---is added to the nucleon momentum distribution.

Using the $\nu T$ package~\cite{ref:vT}, we replace the RFG model by the spectral function (SF) approach~\cite{ref:Omar_RMP}, the implementation of which has been validated through comparisons with electron-scattering data.
In the SF formalism, the interaction between the beam particle and the nucleus is assumed to involve a single nucleon, with the remaining $(A-1)$ nucleons acting as spectators. Under this assumption---called the impulse approximation---the target initial state can be described by the nuclear spectral function $P(\ve p, E)$, giving the probability distribution that removal of a nucleon with momentum $\ve p$ from the nuclear ground state leaves the residual $(A-1)$-nucleon system with excitation energy $E$.

\begin{figure*}
\centering
    \includegraphics[width=0.80\textwidth]{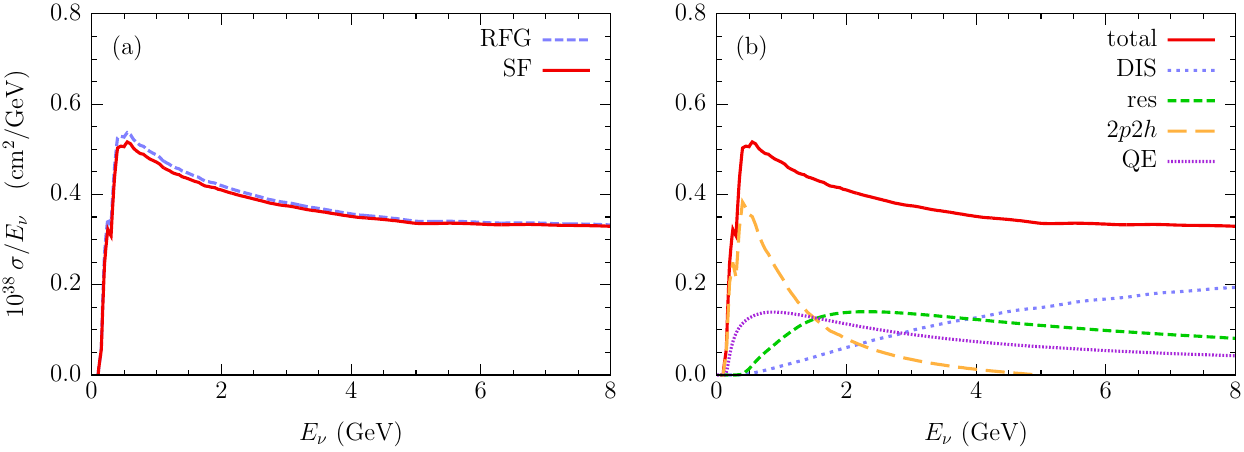}
\caption{\label{fig:xsec_anumu}(color online). Same as in Fig.~\ref{fig:xsec} but for the muon antineutrino.}
\end{figure*}

The realistic SF of carbon, employed in the $\nu T$ package, has been obtained by the authors of Ref.~\cite{ref:Omar_LDA} in the local-density approximation (LDA). The LDA scheme relies on the premise that surface and shell effects do not affect short-range correlations between nucleons in nuclei, and it is supported by the evidence that for $\n p\gtrsim300$~MeV the momentum distribution is largely independent of the mass number $A$, for nuclei with $A\geq4$~\cite{ref:Omar_RMP}.
Therefore, using the LDA, the correlation contribution to the nuclear SF can be obtained from theoretical calculations for uniform nuclear matter at different densities~\cite{ref:Omar_NM,ref:Omar_LDA} and consistently combined with
the shell structure of the nucleus, deduced from experimental $(e,e'p)$ data~\cite{ref:Saclay_C, ref:Dutta}.
We stress that the carbon SF of Ref.~\cite{ref:Omar_LDA} has proven successful in describing electron-scattering data in various kinematical setups, see e.g. Ref.~\cite{ref:FSI}, and the momentum distribution obtained from it is consistent with the one extracted from $(e,e'p)$ data at large missing energy and momentum~\cite{ref:Rohe}.

The effect of using the $\nu T$ package on the total inclusive CC cross section for muon neutrinos in \GENIE{} is presented in Fig.~\ref{fig:xsec_inclusive}. On the one hand, the SF calculation of the QE cross section improves the agreement with the experimental data at low energies reported by the SciBooNE Collaboration~\cite{ref:SciBooNE_inclusive}. On the other hand, it does not alter the good agreement at the kinematics of the NOMAD experiment~\cite{ref:NOMAD_inclusive}, where the inclusive cross section is dominated by the contributions of DIS processes and resonant pion production; see Fig.~\ref{fig:xsec_breakup}.
Note that in the case of the NOMAD data, the error bars represent $\Delta(\sigma/E_\nu)$, whereas for the SciBooNE points, they correspond to $(\Delta\sigma)/E_\nu$~\cite{ref:SciBooNE_inclusive}.

While the calculations do not seem to reproduce the SciBooNE point at the average neutrino energy of 2.47 GeV, it is important to note that its energy range extends over the whole high-energy tail of the flux and does not end at 3.5 GeV.

The recent results of the T2K experiment~\cite{ref:T2K_CC_numu_xsec_C,ref:T2K_CC_numu_xsec_CH} suggest that in the $\sim$1 GeV region uncertainties of the inclusive cross section may be sizable, of the order of 20\%, and that the contribution of the $2p2h$ processes may be smaller than that deduced from MiniBooNE. Should the $2p2h$ estimate be significantly reduced in \GENIE{}, the calculated cross sections would be in very good agreement with the T2K data.

For the sake of completeness, in Fig.~\ref{fig:xsec_anumu}, we show the total inclusive CC cross section for muon antineutrinos, for which experimental data are not available. It clearly appears that the $2p2h$ contribution---which in \GENIE{}~\cite{ref:GENIE_2p} is assumed to be the same for neutrinos and antineutrinos---plays a much more important role in the latter case.

As this article is focused on the effects of realistic detection capabilities on energy reconstruction, we leave comparisons to exclusive cross sections for future studies of nuclear-model uncertainties of the results presented here. Being able to reproduce the inclusive cross sections---of minimal uncertainties---the employed description of neutrino interactions can be deemed suitable for the purpose of this analysis.

\begin{figure}
\centering
    \includegraphics[width=0.80\columnwidth]{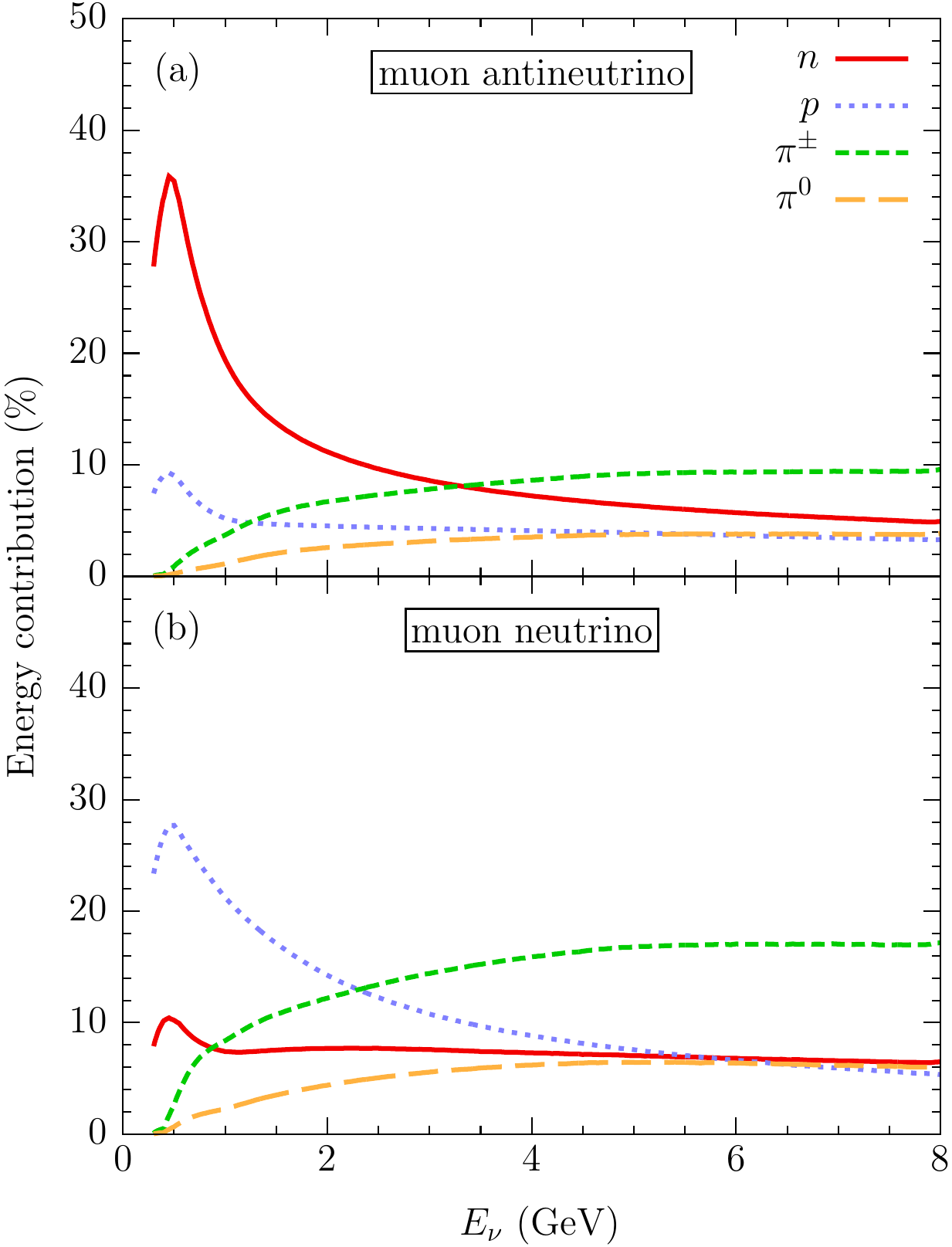}
    \subfigure{\label{fig:energyContribs_anu}}
    \subfigure{\label{fig:energyContribs_nu}}
\caption{\label{fig:energyContribs}(color online). Fraction of the (a) $\bar\nu_\mu$ and (b) $\nu_\mu$ energy converted into the kinetic energies of neutrons (solid lines) and protons (dotted lines) and the total energies of charged (short dashed lines) and neutral (long dashed lines) pions.}
\end{figure}

In the context of the calorimetric energy reconstruction, it is important to estimate how different species of hadrons contribute to the final-state energy. In Fig.~\ref{fig:energyContribs}, we present the obtained fraction of the $\bar\nu_\mu$ and $\nu_\mu$ energy carried out by neutrons, protons, and pions.

While at energies below $\sim$1 GeV the neutron results for neutrinos and antineutrinos clearly differ, this is not the case when DIS and resonant pion production become the dominant reaction mechanisms. Having in mind the numerical uncertainties---estimated not to exceed 1\%---we conclude that at $E_\nu\gtrsim1.5$ GeV neutrons contribute less than 15\% of the (anti)neutrino energy.

As can be expected from QE interactions, the contribution of protons in neutrino interactions resembles that of neutrons in antineutrino scattering and vice versa.
For antineutrinos, the knocked-out protons carry out less than 10\% of the initial energy in the whole considered region. However, for neutrinos, the protons contribution is much more sizable and reduces to $\leq10$\% for energies exceeding $\sim$3.35 GeV.

Although the total energies of $\pi^0$'s enter the calorimetric energy reconstruction, in the considered kinematic range, their contributions are smaller than that of neutrons. On the other hand, in the region dominated by resonant and nonresonant pion production, charged pions give the largest contribution to the final-state hadronic energy.

We observe that at energies above 0.5 GeV the analogical results for electron neutrinos and antineutrinos do not differ significantly from those presented in Fig.~\ref{fig:energyContribs}.

\section{Considered detector effects}
\label{sec:scenarios}

In our analysis, two extreme sets of assumptions are considered regarding the detector performance:
\begin{itemize}
\item[(i)] Perfect reconstruction.\\
With the exception of neutrons, all produced particles are observed, and their measured energies are equal to the true ones.
\item[(ii)] Realistic setup.\\
The measured energies and angles are smeared with respect to the true ones by a finite detector resolution. The detection efficiencies and thresholds are taken into account. Neutrons are assumed to escape detection.
\end{itemize}

The detection thresholds applied in our calculations correspond to the measured kinetic energy of 20 MeV for mesons and 40 MeV for protons. For comparison, the NOMAD and MiniBooNE experiments were able to detect protons of kinetic energy above $\sim$50 MeV~\cite{ref:NOMAD}, and the $\sim$40-MeV threshold is expected for future liquid-argon detectors~\cite{Barger:2007yw}.

The efficiencies are treated as energy independent, for the sake of simplicity, and set to 60\% for $\pi^0$'s, 80\% for other mesons, and 50\% for protons. Those values can be considered rather optimistic, compared to the efficiencies achieved in existing detectors~\cite{ref:NOMAD,ref:efficiency_pi0_ND280,ref:efficiency_piC_MINERvA}. We assume that produced charged leptons are always detected.

Accounting for the effect of finite detector resolution on an observable $x_\textrm{meas}$, we smear it according to the Gaussian distribution centered at the true value $x_\textrm{true}$,
\[
f(x_\textrm{meas})=\frac1{\sqrt{2\pi}\sigma(x_\textrm{true})}\exp\left[-\frac12\left(\frac{x_\textrm{meas}-x_\textrm{true}}{\sigma(x_\textrm{true})}\right)^2\right].
\]

In the case of the muon, we apply this procedure to the momentum and production angle, using
\begin{equation}\label{eq:muonRes}
\sigma(\n{k_\mu})=0.02\n{k_\mu}\quad\textrm{ and }\quad\sigma(\theta)=0.7^\circ,
\end{equation}
as in the MINERvA experiment~\cite{ref:MINERvA}. For the electron, the realistic resolutions~\cite{ref:MiniB_nue,ref:NOvA}
\begin{equation}\label{eq:eRes}
\sigma(E_e)=0.10E_e\quad\textrm{ and }\quad\sigma(\theta)=2.8^\circ
\end{equation}
are used.

For other particles, the only smeared quantity is the energy.
To the electromagnetic showers produced by $\pi^0$ decays, we employ the energy resolution given by
\begin{equation}\label{eq:pi0Res}
\frac{\sigma(E_{\pi^0})}{E_{\pi^0}}=\max\left\{\frac{a_{\pi^0}}{\sqrt{E_{\pi^0}}},\, \frac{b_{\pi^0}}{E_{\pi^0}}\right\}
\end{equation}
with $a_{\pi^0}=0.107$ and $b_{\pi^0}=0.02$, while for other hadrons,
\begin{equation}\label{eq:hadrRes}
\frac{\sigma(E_h)}{E_h}=\max\left\{\frac{a_h}{\sqrt{E_h}},\, b_h\right\}
\end{equation}
is used with $a_h=0.145$ and $b_h=0.067$. The values of $\sigma$ and the energies appearing in Eqs.~\eqref{eq:pi0Res} and \eqref{eq:hadrRes} are expressed in units of GeV. Note that the hadron-energy resolutions applied in our analysis can be considered optimistic, as they are to be compared to
\[
\frac{\sigma(E_{\pi^0})}{E_{\pi^0}}=2\sqrt{\frac{a_{\pi^0}^2}{E_{\pi^0}}+ \frac{b_{\pi^0}^2}{E_{\pi^0}^2}}
\quad\textrm{and}\quad
\frac{\sigma(E_h)}{E_h}=2\sqrt{\frac{a_h^2}{E_h}+b_h^2}
\]
achieved in the MINOS~\cite{ref:MINOS} and MINERvA~\cite{ref:MINERvA} experiments. Note that the assumed $\pi^+$ energy resolution is also optimistic when compared with the expectation for future liquid-argon detectors from Ref.~\cite{Stahl:2012exa}.

It is important to emphasize that in the context of the kinematic method of energy reconstruction our assumptions are conservative. We employ realistic detector resolutions for muons and make minimal use of information on observed hadrons. Disregarding additional protons in the final state, we reconstruct every event as if it involved single-nucleon knockout.
All mesons in the final state are assumed to originate from a single resonance of the invariant mass $M_\Delta=1.232$ GeV, and their angular distributions are not taken into account; compare to Eq.~\eqref{eq:kinRecEnergy_multitrack}.
On the other hand, detector effects entering the calorimetric method are treated in a rather optimistic way, compared to those in existing detectors.

Note that knocked-out neutrons are assumed to be undetected both in the realistic and perfect scenarios, owing to difficulties of their reconstruction in neutrino events. Traveling some distance from the primary interaction vertex before scattering, neutrons are currently problematic to associate with the neutrino event, and they typically deposit only part of their energy in the detector. However, this assumption may need to be revisited in the future, when an ongoing experimental program~\cite{Berns:2013usa,Liu:2015fiy} brings important progress in the understanding of the detector response to neutrons.

\section{Migration matrices}
\label{sec:migrationMatrices}

In our calculations, detector effects are taken into account in the migration matrices, $\mathcal{M}^{X}_{ij}$,
the columns of which are the probability distribution functions (PDFs) for a neutrino interaction $X$ at the true energy in the $i$th bin to be reconstructed with an energy in the $j$th bin. The observed event distribution can then be calculated as
\[ N^\textrm{tot}_{i} = \sum_{X} \sum_{j} \mathcal{M}^{X}_{ij} N^{X}_j ,
\]
where $X$ runs over the four types of interactions considered (DIS, res, $2p2h$, and QE), $i$ and $j$ refer to the energy bins, and $N^X_j$
stands for the number of $X$ events in the bin $j$ computed without detector effects. Note that, barring the boundary effects, the smearing produced by the
migration matrices does not have any impact on the
total number of events.

To facilitate the reproduction of our results, we provide the complete set of migration matrices, calculated for the energies up to 8 GeV, using 0.1~GeV bins, and the cross sections employed in our analysis~\cite{matricesRef}.

\begin{figure}
\begin{center}
	\includegraphics[width=0.80\columnwidth]{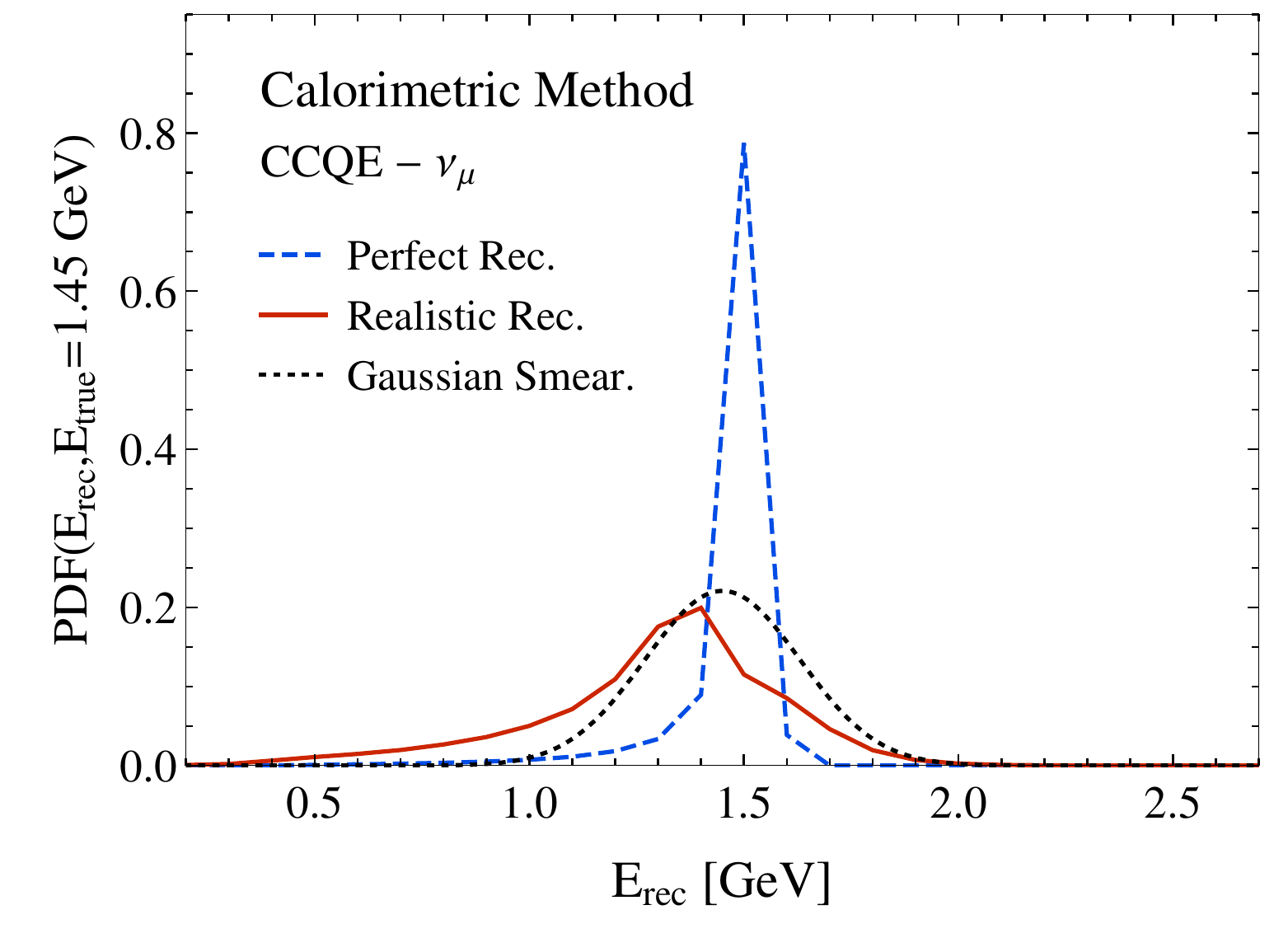}
	\includegraphics[width=0.80\columnwidth]{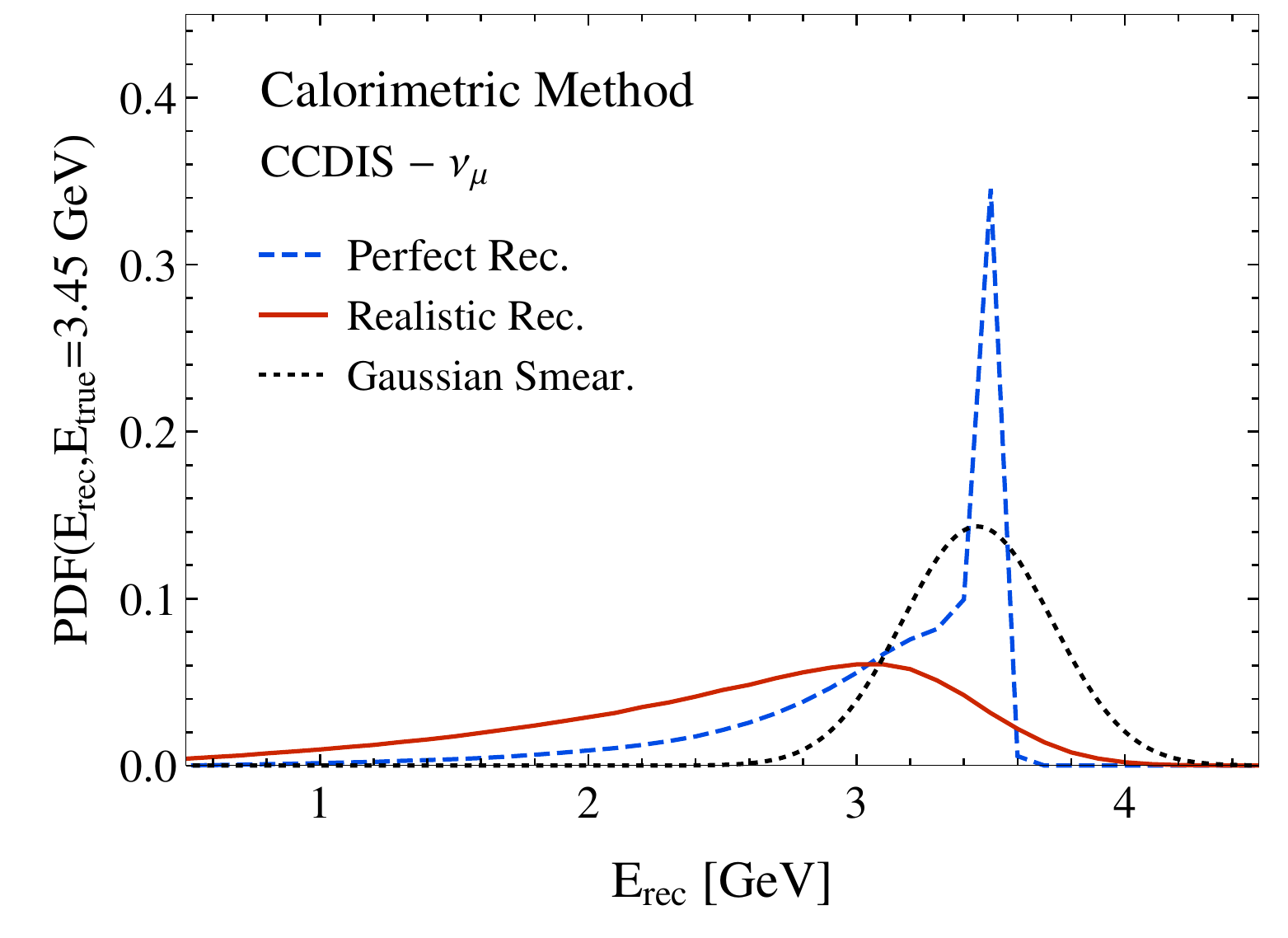}
\caption{(color online). Probability distribution functions for a $\nu_\mu$ event
  of the energy $E_\textrm{true}$ to be reconstructed at an energy $E_\textrm{rec}$. The results obtained for a QE event at $E_\textrm{true} = 1.45$ GeV (top) and a DIS event at $E_\textrm{true} = 3.45$ GeV (bottom) show the effects of different assumptions on the detector performance; see the text
  for details. }
\label{fig:pdfs}
\end{center}
\end{figure}

\begin{figure}
\begin{center}
	\includegraphics[width=0.80\columnwidth]{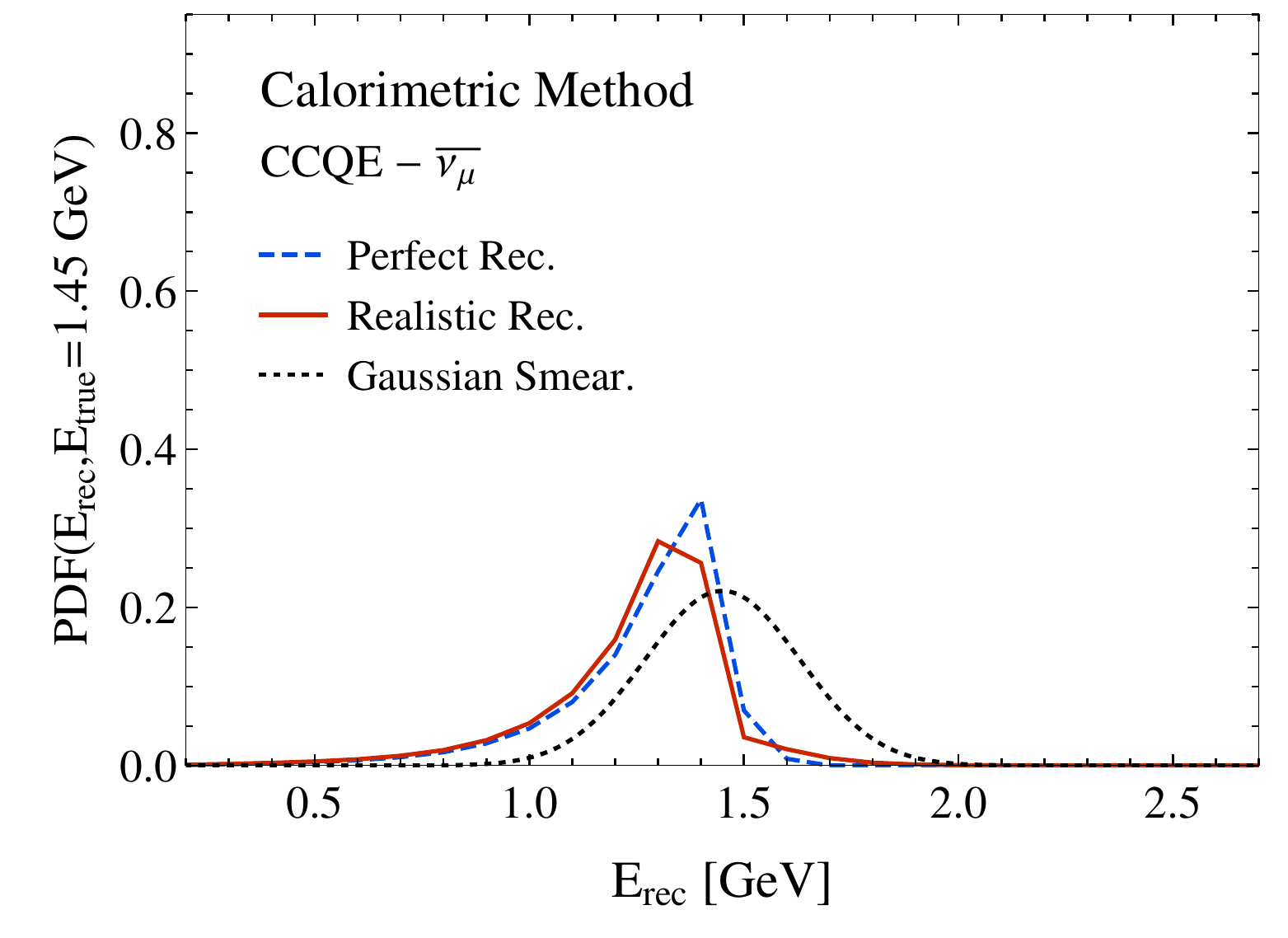}
	\includegraphics[width=0.80\columnwidth]{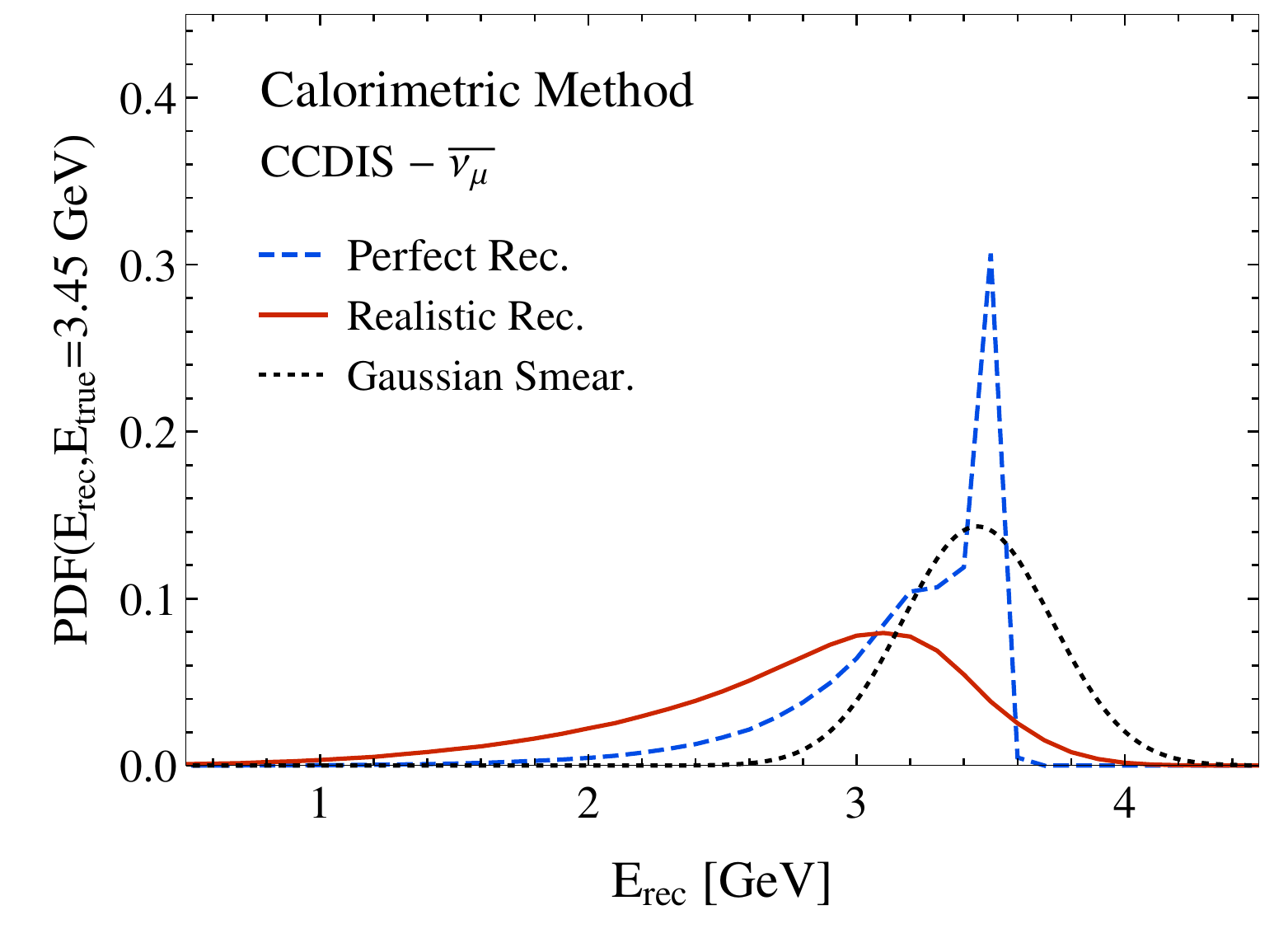}
\caption{(color online). Same as in Fig.~\ref{fig:pdfs} but for a $\bar\nu_\mu$ event. }
\label{fig:pdfs_anu}
\end{center}
\end{figure}

In an ideal detector, all migration matrices would be the unit matrices. In a real experiment, the reconstructed energy may depend on the reconstruction procedure, as well as on the mechanism of interaction. Imperfect detection capabilities---energy resolutions, efficiencies, and thresholds for particle detection---affect the probability for a neutrino event to be reconstructed in the correct energy bin. In particular, finite energy resolutions smear the measured energies, while imperfect efficiencies and finite thresholds result in an energy partially carried away by undetected particles. As a consequence, PDFs have finite widths, they are asymmetric, with a broader tail toward the lower energies, and their mean values are lowered with respect to the true neutrino energies.

Those features are illustrated in Figs.~\ref{fig:pdfs} and~\ref{fig:pdfs_anu} for the QE (DIS) mechanism of interaction and the true neutrino and antineutrino energy of 1.45 GeV (3.45 GeV) reconstructed using the calorimetric method. Comparing the PDFs obtained for the perfect and realistic reconstructions, defined in Sec.~\ref{sec:scenarios}, one can observe how sizable the effect of realistic detector capabilities is. For reference, we also show the Gaussian distribution with the standard deviation $\sigma(E_\nu) = 0.15\sqrt{E_\nu}$, with $\sigma$ and $E_\nu$ in units of GeV, typically applied to account for detector effects in phenomenological studies devoted to liquid-argon detectors~\cite{Barger:2007yw,ref:LBNE,Agarwalla:2012bv,Agarwalla:2013qfa,Agarwalla:2011hh}.

The differences between the neutrino and antineutrino PDFs can be traced back to a twofold reason: (i) different contributions of neutrons to the final-state energy, as shown in Fig.~\ref{fig:energyContribs}, and (ii) the typical energy transfer being lower in antineutrino interactions, owing to the destructive interference of the response functions.

Even for the perfect-reconstruction scenario, the PDFs are asymmetric, as a consequence of pion absorption in the nuclear medium and the energy carried out by neutrons, assumed to escape detection. When the realistic detector effects are accounted for, the PDFs clearly broaden due to the employed energy resolutions, and their modes shift toward lower energies. The latter effect is particularly large for DIS events, in which the muon contribution to the final-state energy is typically smaller than in QE scattering, and the role of the efficiencies is larger.

\begin{figure}
\begin{center}
	\includegraphics[width=0.80\columnwidth]{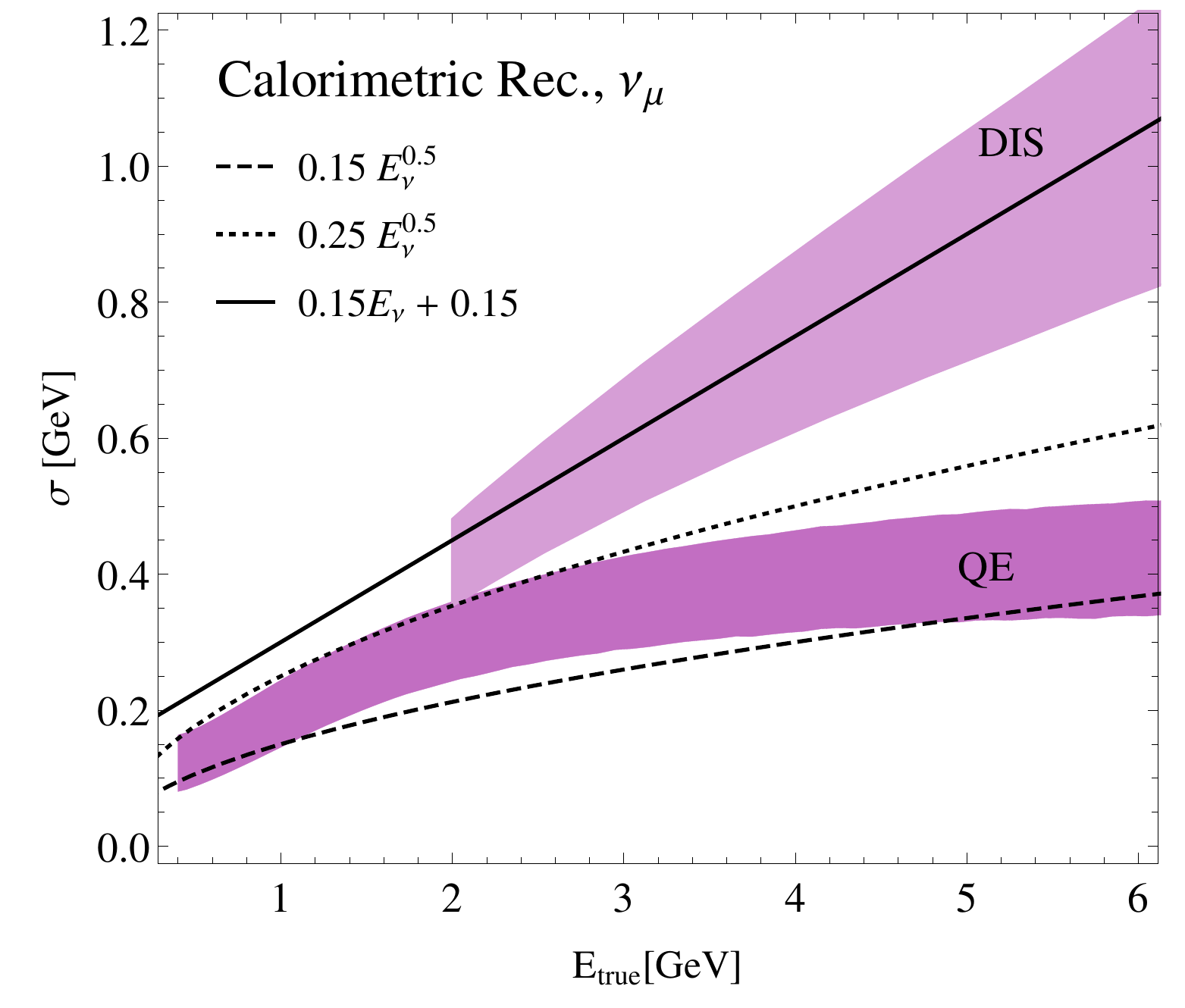}
	\includegraphics[width=0.80\columnwidth]{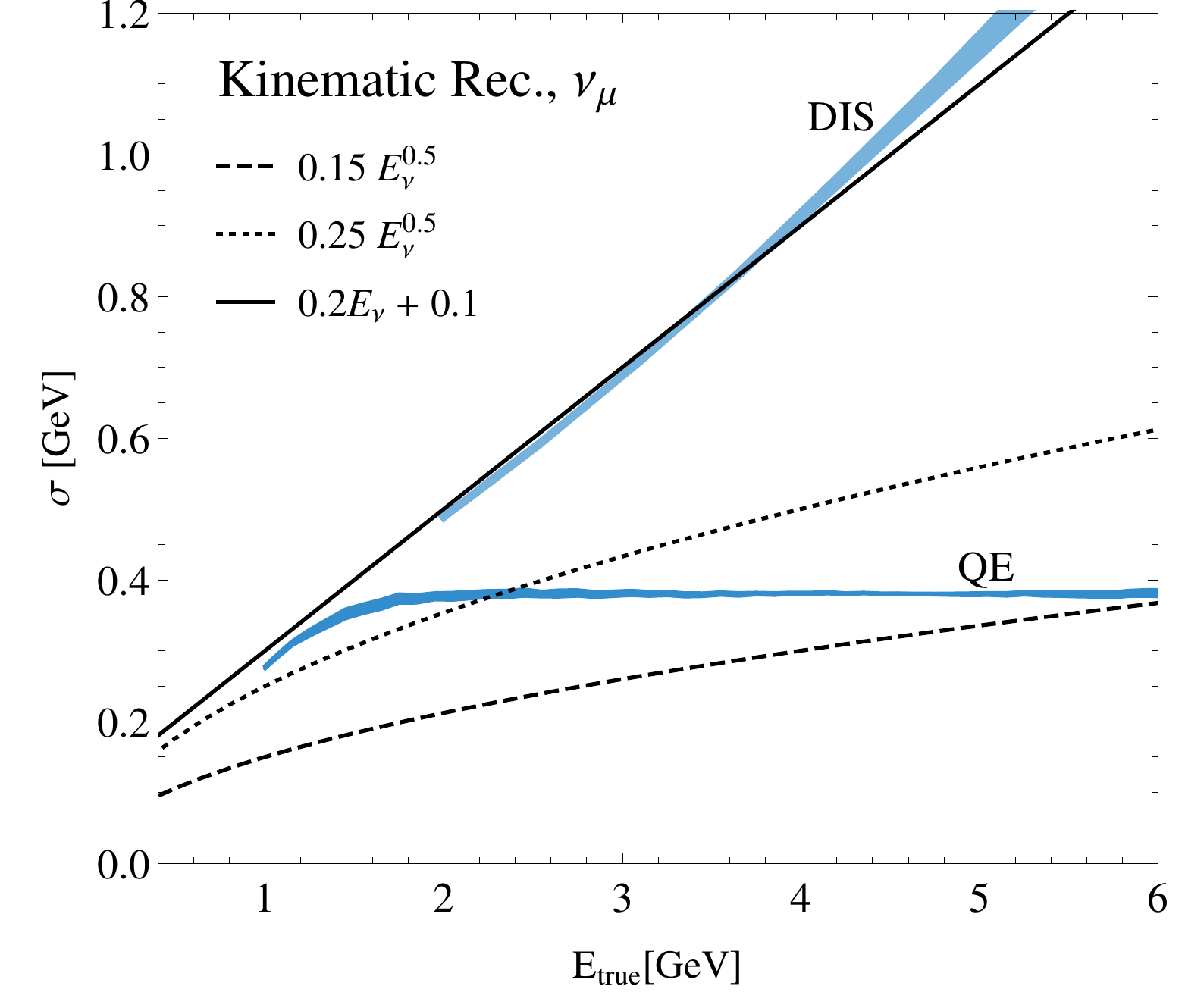}
\caption{(color online). Effective energy resolution as a function
  of the true energy for muon neutrinos. The results of our Monte Carlo simulations for QE (DIS) events are shown as lower (upper) bands. For each band, the upper (lower) edge corresponds to the realistic (perfect) detection capabilities, defined in Sec.~\ref{sec:scenarios}.
  For comparison, a few smearing functions frequently used in  phenomenological oscillation studies are also shown. }
\label{fig:effRes}
\end{center}
\end{figure}

\begin{figure}
\begin{center}
	\includegraphics[width=0.80\columnwidth]{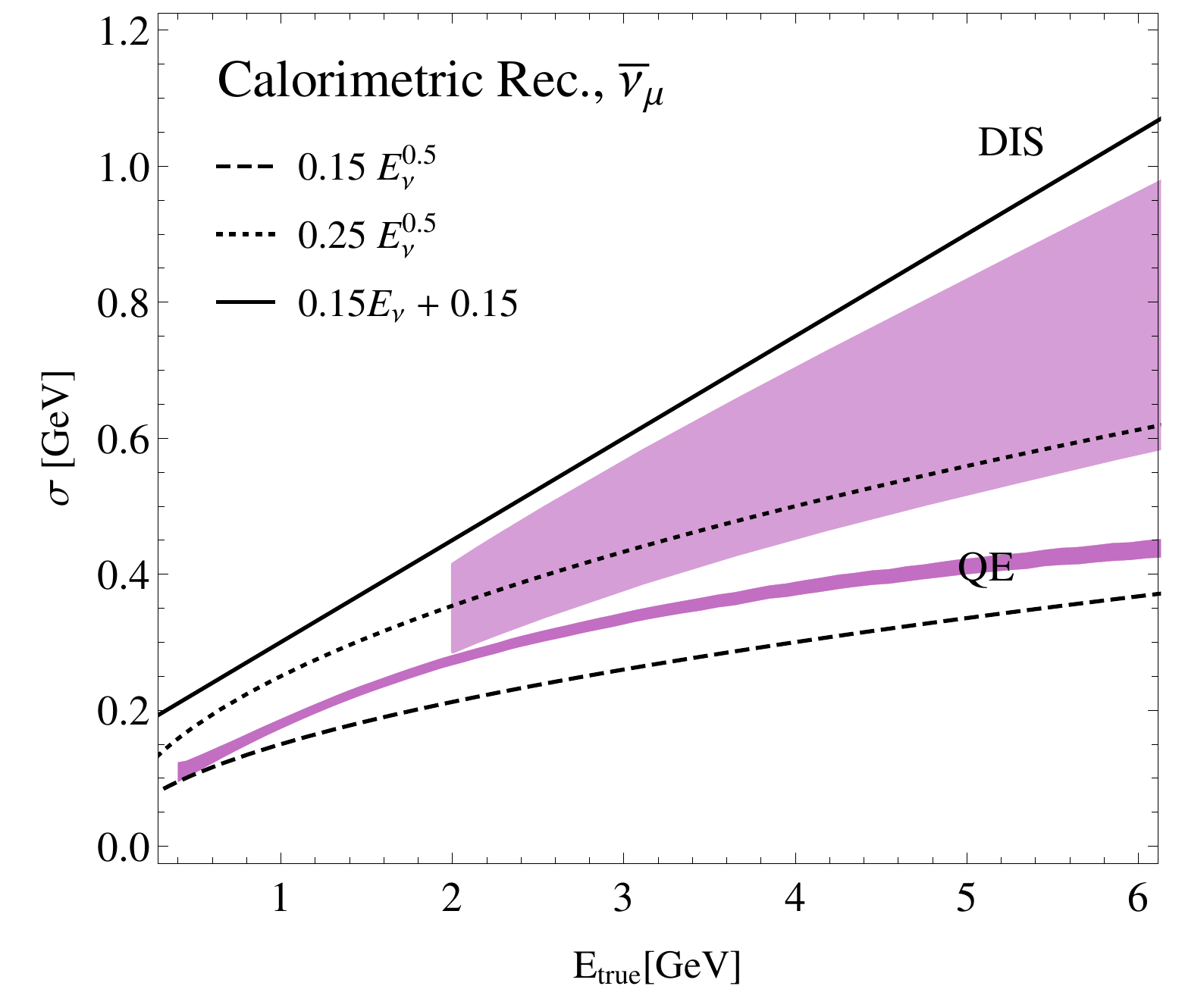}
	\includegraphics[width=0.80\columnwidth]{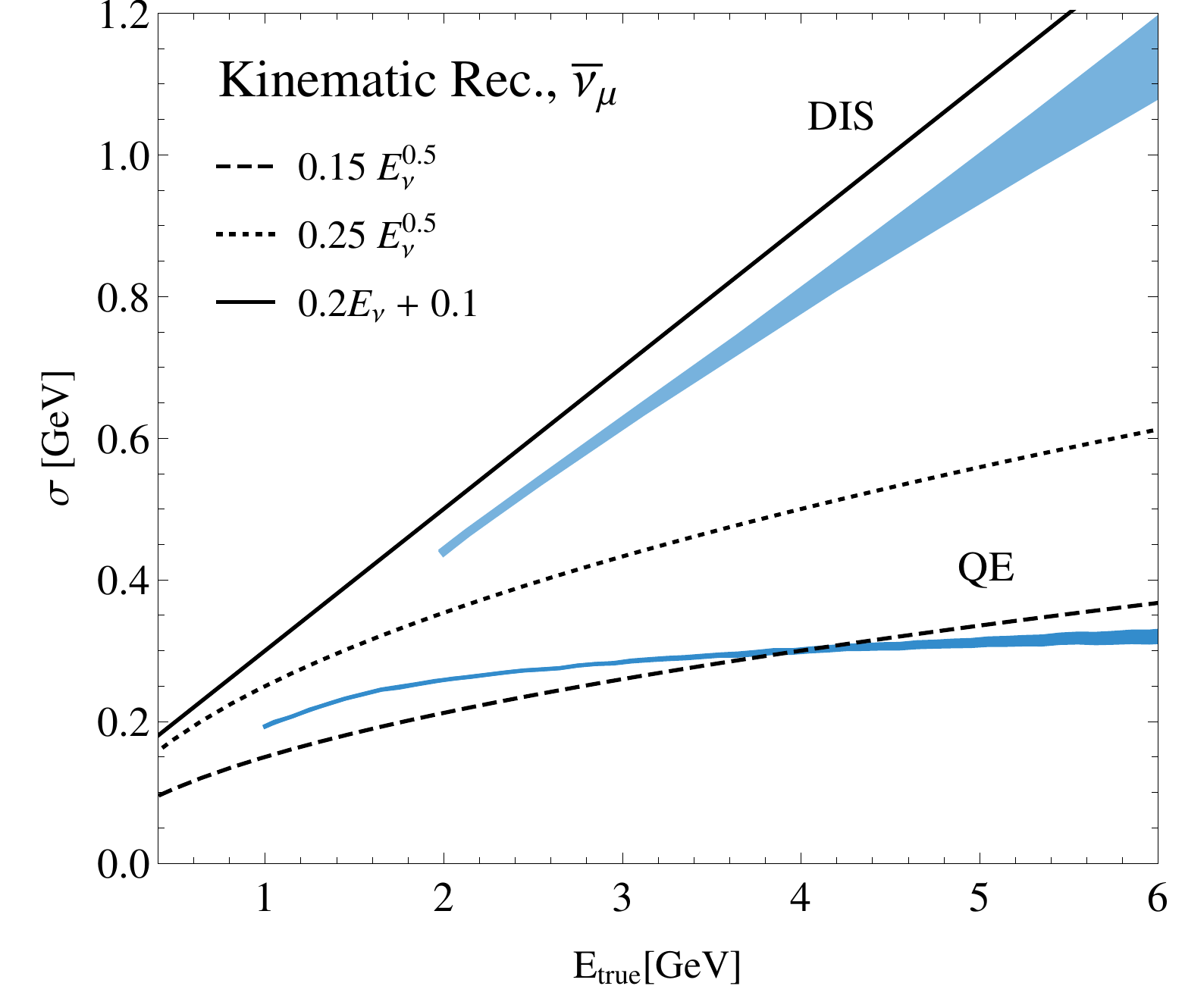}
\caption{(color online). Same as in Fig.~\ref{fig:effRes} but for $\bar\nu_\mu$'s. }
\label{fig:effRes_anu}
\end{center}
\end{figure}

To make contact with existing phenomenological studies of neutrino and antineutrino oscillations,
in Figs.~\ref{fig:effRes} and \ref{fig:effRes_anu}, we compare a few simple functions typically employed as an effective energy resolution with our estimates based on the Monte Carlo simulations, for both the calorimetric (upper panels) and kinematic (lower panels) reconstruction methods. Our calculations of the PDF's standard deviations are presented as bands spanning the values between the results for the perfect-reconstruction scenario (lower edge) and those for the realistic scenario (upper edge). The estimates for QE and DIS events are shown separately, as the lower (darker) and upper (lighter) bands. Because the energy resolutions applied in the realistic scenario exceed those in existing experiments roughly two times, they can be considered optimistic. On the other hand, an effective energy resolution higher than our result for the perfect-reconstruction scenario does not seem to be possible to achieve without the reconstruction of neutrons.

Note that for liquid-argon experiments (such as DUNE or LBNO) or totally active scintillator
detectors (e.g., NO$\nu$A), the effective energy resolution is typically assumed to be  $\sigma(E_\nu)=\mathcal{O}(0.05\textrm{--}0.2) \sqrt{E_\nu}$, in units of GeV; see, for instance, Refs.~\cite{Barger:2007yw,Patterson:2012zs,ref:LBNE,Agarwalla:2012bv,Agarwalla:2011hh,Agarwalla:2013qfa}.
On the other hand, for experiments based on the Cherenkov technique (such as T2K), the resolution
$\sigma(E_\nu)=\mathcal{O}
(0.08)\sqrt{E_\nu}$ is usually applied in phenomenological studies~\cite{Huber:2009cw,Agarwalla:2013qfa}.

\begin{figure}
\begin{center}
	\includegraphics[width=0.80\columnwidth]{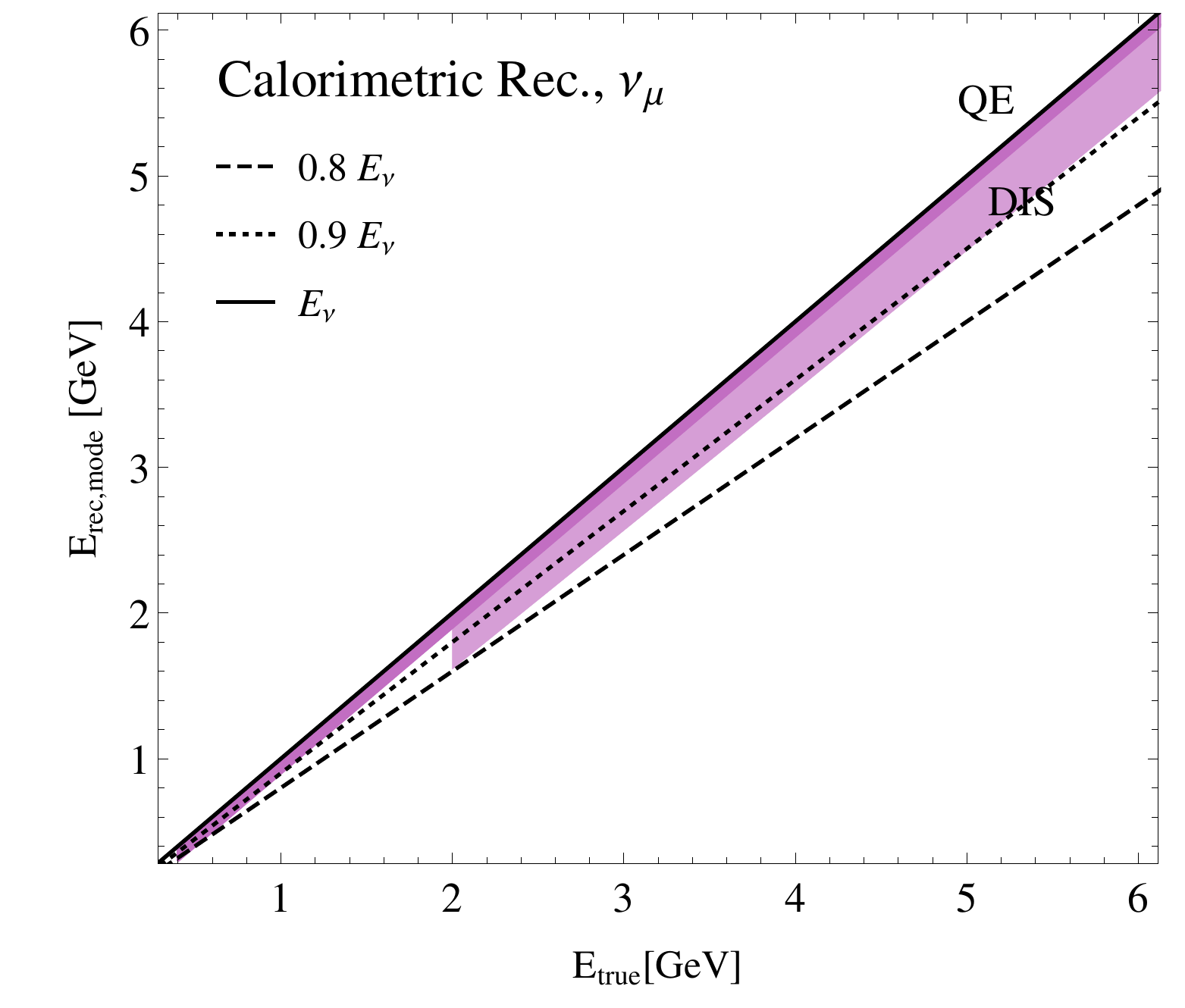}
	\includegraphics[width=0.80\columnwidth]{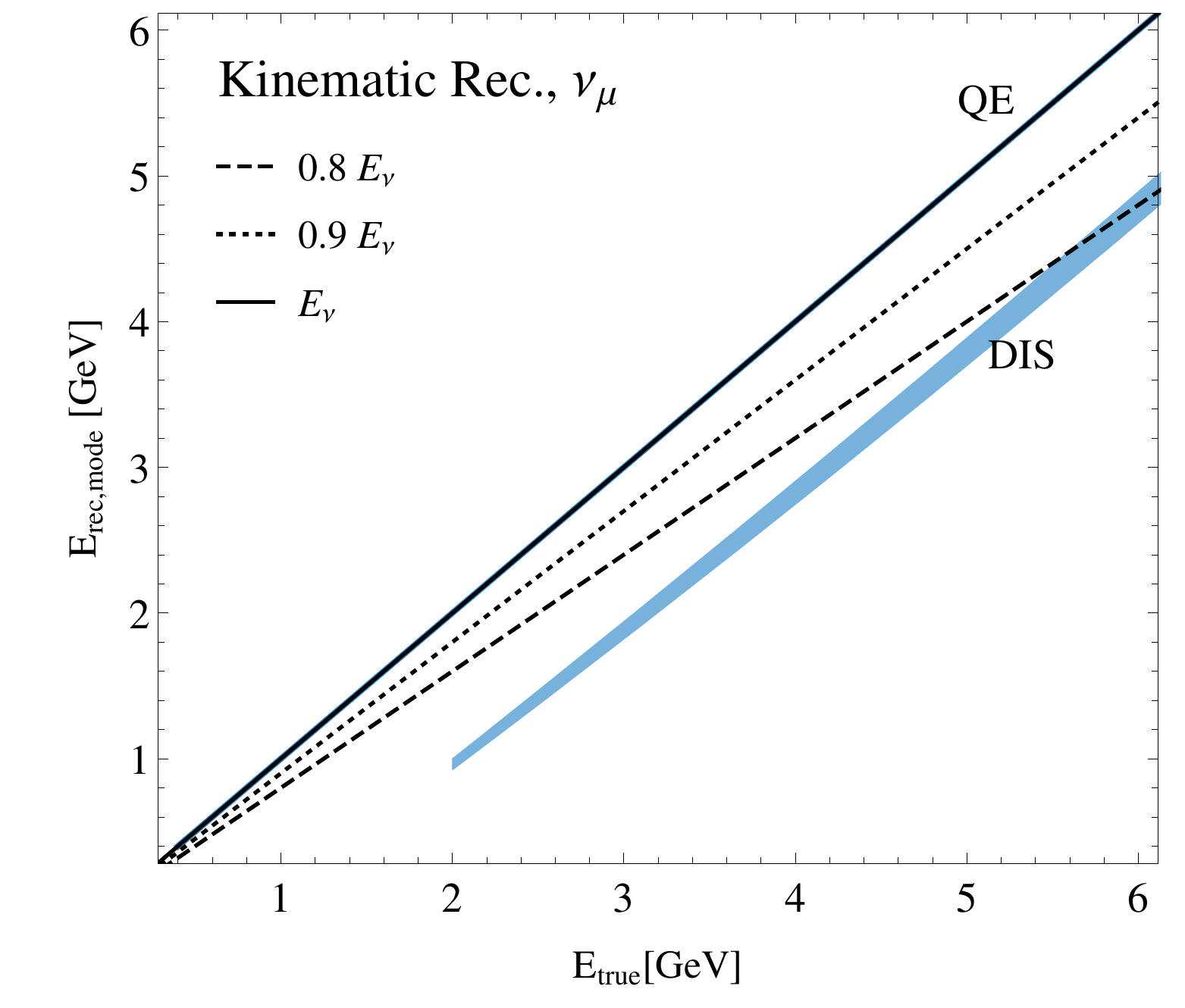}
\caption{(color online). Mode of the reconstructed-energy distributions as a function of the true energy calculated for muon neutrinos. The bands show our Monte Carlo results, with the lower (upper) edge obtained assuming the realistic (perfect-reconstruction) scenario. The darker (lighter) bands present the results for QE (DIS) events. For reference, the lines corresponding to the true value and its underestimation by 10 and 20\% are also shown. }
\label{fig:effMean}
\end{center}
\end{figure}

\begin{figure}
\begin{center}
	\includegraphics[width=0.80\columnwidth]{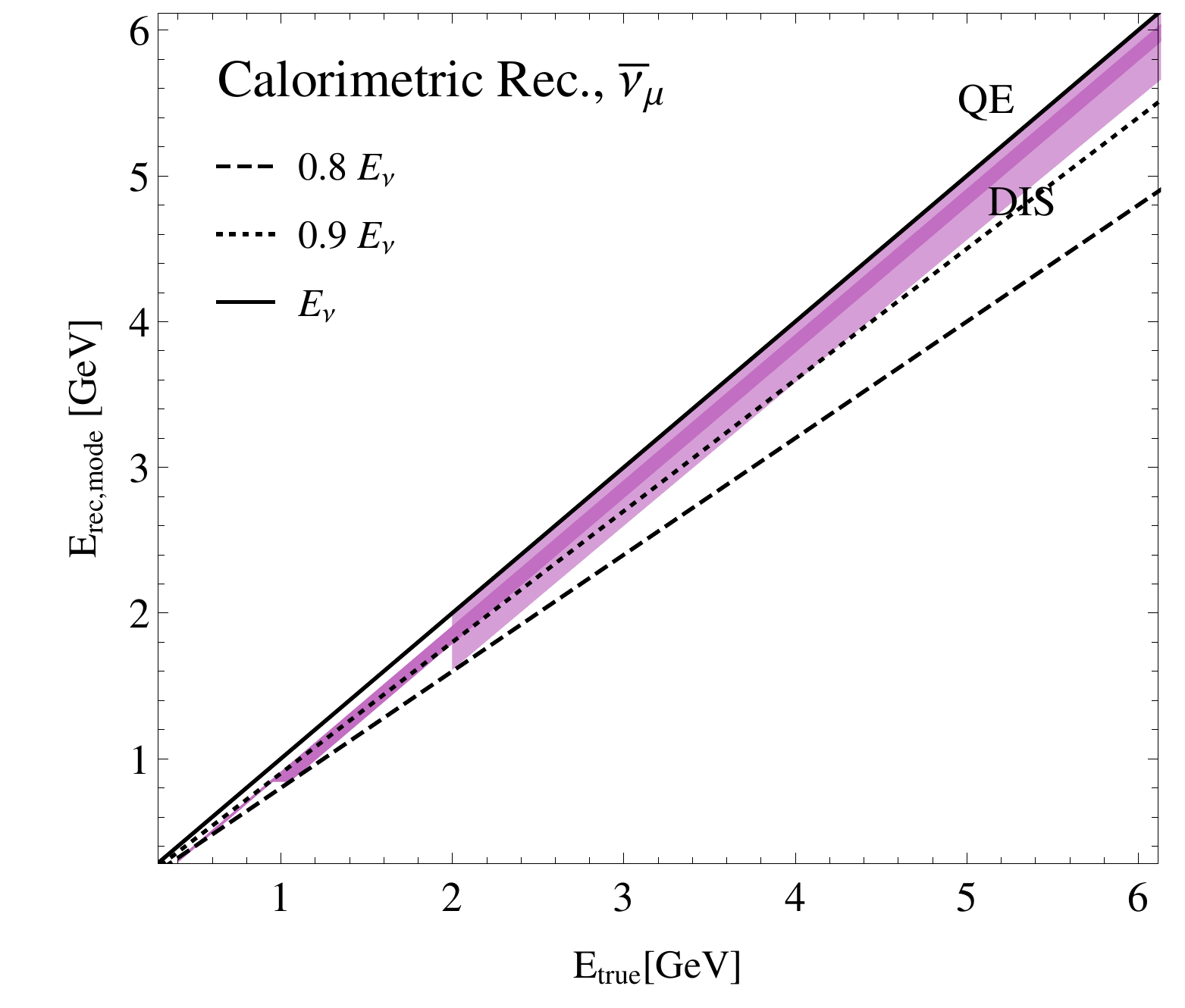}
	\includegraphics[width=0.80\columnwidth]{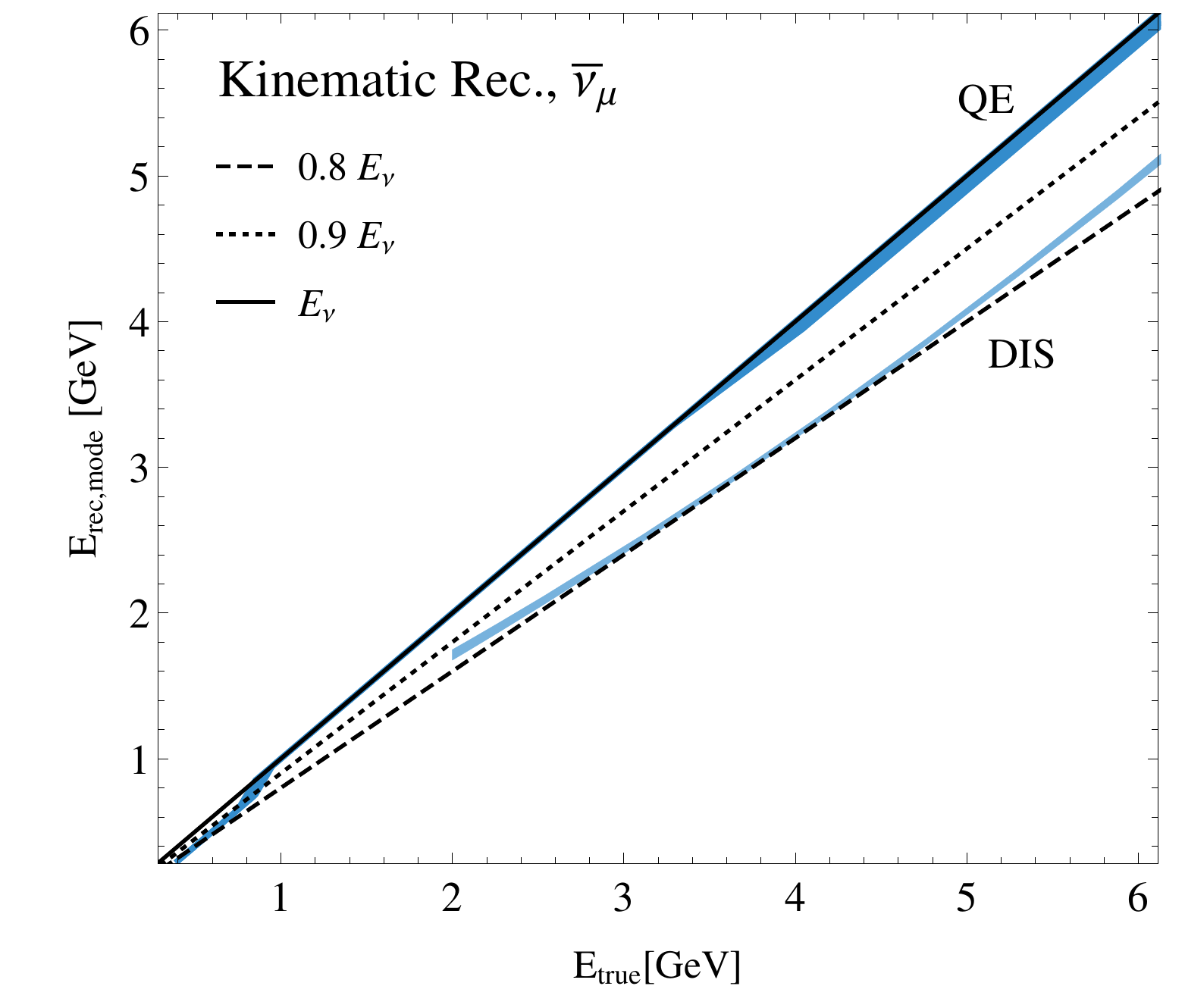}
\caption{(color online). Same as in Fig.~\ref{fig:effMean} but for $\bar\nu_\mu$'s. }
\label{fig:effMean_anu}
\end{center}
\end{figure}

Finally, in Figs.~\ref{fig:effMean} and \ref{fig:effMean_anu}, we show how the modes of the reconstructed-energy distributions depend on the true value of energy, comparing the calorimetric and kinematic reconstruction methods. The bands represent our Monte Carlo results for QE and DIS events, the lower (upper) edge of which corresponds the realistic (perfect-reconstruction) scenario.

In the calorimetric method, the modes for QE scattering are in a good agreement with the true energy, and the expected presence of neutrons in the final state of antineutrino events introduces only a small effect. While for DIS, the agreement is somewhat reduced---especially when detector effects are taken into account---the modes do not differ from the true energy by more than 20\%.

For the kinematic method, in QE interactions, both the neutrino and antineutrino modes are in excellent agreement with the true energy. However, this is not the case for DIS events. For antineutrinos, the discrepancy between the DIS mode and the true energy exceeds 10\% and is typically close to 20\%. For neutrinos, the mode underestimates the true energy by $\sim$1 GeV. This behavior can be traced back to the simplicity of our kinematic analysis, assuming that all events containing at least one pion are produced through excitation of a resonance of the invariant mass $M_\Delta=1.232$ GeV; see Eq.~\eqref{eq:kinRecEnergy_W}. Should an experiment be able to reconstruct pions angles with high angular resolution, more refined treatment~\eqref{eq:kinRecEnergy_multitrack} would be possible.

The bands in Figs.~\ref{fig:effRes}--\ref{fig:effMean_anu} show the differences between the results for the realistic and perfect-reconstruction scenarios. Their widths can be used as a measure of the sensitivity to detector effects,
which turns out to be larger for the calorimetric migration matrices than for the kinematic ones.
However, they do not represent uncertainties entering any actual experiment. In practical situations, detection capabilities and their uncertainties are estimated in test-beam exposures. In Sec.~\ref{sec:oscillationAnalysis}, we are going to analyze the importance of an accurate estimate of these uncertainties in the context of an oscillation analysis.

\section{Oscillation analysis}
\label{sec:oscillationAnalysis}

The event number expected in a neutrino-energy bin, without considering detector effects, can be computed as
\begin{equation}
\label{eq:Nev}
N^i_{\beta}  = \int_{E_\nu^i}^{E_\nu^i+\Delta E_\nu}  d E_\nu\: \sigma_{\nu_\beta}(E_\nu)\:P_{\alpha \beta}\left(\{\Theta\}, E_\nu\right)\:\phi_{\nu_\alpha}(E_\nu),
\end{equation}
where the indices $\alpha$ and $\beta$ label the initial and final
neutrino flavors, respectively; $\Delta E_\nu$ is the energy-bin size;
$\sigma_{\nu_\beta}$ denotes the ${\nu_\beta}$ cross section; and
$\phi_{\nu_\alpha}$ is the unoscillated $ \nu_\alpha$ flux.
 The $\nu_\alpha \rightarrow \nu_\beta$ oscillation
probability, $P_{\alpha \beta}$, depends on a set of oscillation
parameters $\left\{\Theta\right\}$ as well as on the neutrino energy
$E_\nu$. In the $\nu_\mu \rightarrow \nu_\mu$ oscillation channel, the oscillation probability can be approximated by
\[
P_{\mu\mu} \simeq 1 - \sin^2\theta_{\mu\mu}\sin^2\left(\frac{ \Delta m^2_{\mu\mu}L}{4E_\nu}\right),
\]
where
\[
\sin^2\theta_{\mu\mu} = 4\cos^2\theta_{13}\sin^2\theta_{23}(1 - \cos^2\theta_{13}\sin^2\theta_{23}),\]
and $\Delta m^2_{\mu\mu}$ is the weighted average of $\Delta m_{31}^2$ and $\Delta m_{32}^2$~\cite{Nunokawa:2005nx} that can be expressed as
\[
\Delta m^2_{\mu\mu}=\Delta m^2_{31} + \mathcal{O}(\Delta m^2_{21}).
\]

Should the experimental information on the $E_\nu$ distribution of
events be available, the values of the oscillation parameters
$\{\Theta\}$ could be, in principle, directly inferred from the
data\footnote{Note, however, that if more than one parameter is being determined from the
  data, severe degeneracies among the different parameters may take
  place.\cite{Minakata:2001qm,BurguetCastell:2001ez,Barger:2001yr,Donini:2003vz,Fogli:1996pv,Minakata:2013hgk,Coloma:2014kca}} However,
modern neutrino beams are produced as tertiary products, originating
predominantly from the decay of pions and kaons produced in
interactions of protons impinging on a target. Therefore, neutrinos
are not monoenergetic, and---for a given event---the incident neutrino
energy has to be reconstructed from the kinematics of the
particles in the final state. Systematic uncertainties of this
procedure inevitably depend on the capabilities of the employed detector,
as well as on the neutrino interaction channel, owing to the different
detection efficiencies for the particles involved in the event. Because for
muons the efficiency uncertainty is minimal, its effect on the
reconstructed energy spectrum is neglected in our analysis.

The kinematic method of energy reconstruction---used, for instance, in
Cherenkov detectors---is known to be accurate to $\sim$100--150 MeV
for QE events~\cite{ref:Ankowski_NuFact2014}. However, as discussed in
Sec.~\ref{sec:energyReconstruction}, this is not the case for the
events of QE topology containing undetected hadrons. For example, a
single undetected pion typically spoils the reconstructed energy by
$\sim$300--350 MeV~\cite{ref:Leitner}. For pion-production events,
playing an important role at higher energies, either the accuracy is
reduced [see Eq.~\eqref{eq:kinRecEnergy_W} containing $W^2$, unknown
  on event-by-event basis] or detailed angular information is required
[see Eq.~\eqref{eq:kinRecEnergy_multitrack}].

Phenomenological analyses for wide-band neutrino
beams operating in the multi-GeV energy regime have shown that the sensitivity to neutrino oscillation
parameters of a liquid-argon detector is similar to that of a Cherenkov detector of the mass $\sim$3--6 times larger, because of the lower efficiency of the latter for non-QE events~\cite{Barger:2007yw,Akiri:2011dv,Huber:2010dx,Coloma:2012ut,Coloma:2012ma}. It
has been argued that the reason lies in the imaging capabilities of
the liquid-argon technology, able to identify protons in addition to
lepton tracks~\cite{Barger:2007yw}, as opposed to Cherenkov detectors
in which the information on CC events comes predominantly from the charged
leptons. Nevertheless, in such comparisons, the neutrino energy
resolution at liquid-argon detectors was always assumed to be
extremely good, in the range of $\sigma (E_\nu) = 0.15
\sqrt{E_\nu}$. Should this be affected by detector effects, or by the
energy carried away by unobserved particles, such conclusions may have
to be reexamined.

In this work, we compare the kinematic and calorimetric methods of
neutrino energy reconstruction in the context of a $\nu_\mu$
disappearance experiment.  Four types of CC neutrino interactions are
considered---resonant and nonresonant pion production, two-nucleon
knockout, and quasielastic scattering---and modeled according to
\GENIE{} with the $\nu T$ package, as described in
Sec.~\ref{sec:GENIE}. The coherent channel has not been taken into
account because of its negligible contribution to the event
rate~\cite{ref:Formaggio}. For each interaction type and both
reconstruction methods, we have calculated corresponding migration
matrices $\mathcal{M}^X_{ij}$ with and without detector effects; see
Secs.~\ref{sec:scenarios} and~\ref{sec:migrationMatrices}.

In the $\nu_\mu \rightarrow \nu_\mu$ oscillation channel, the main
background comes from neutral current events misidentified as CC
ones. As it is generally expected to be very low, for simplicity, we
neglect it in our analysis. On the other hand, due to the large event
statistics, systematic uncertainties are expected to be relevant. In
their treatment, we follow that of Refs.~\cite{ref:Pilar_PRL,
  ref:Pilar_PRD}, with the $\chi^2$ implementation as detailed in the
appendices of Refs.~\cite{ref:Pilar_PRD,ref:Coloma:2012ji}. A 20\%
bin-to-bin uncorrelated systematic uncertainty is assumed, as well as
a 20\% overall normalization uncertainty, which is bin-to-bin
correlated. The pull method is used, adding a Gaussian prior for each
systematic error, and the final $\chi^2 $ profile is obtained after the
minimization over the nuisance parameters. These (rather conservative)
systematic uncertainties are introduced in our analysis in order to
accommodate possible differences in the shape of the expected
event distributions at the detectors, in a similar fashion as was
done in Refs.~\cite{ref:Pilar_PRL, ref:Pilar_PRD}. A near detector is
also considered in the analysis, which helps to constrain the nuisance
parameters during the fit.

The oscillation analysis is performed using a modified
version~\cite{ref:Coloma:2012ji} of \GLoBES{}~\cite{GLB1, GLB2}. The assumed true values of
the oscillation parameters are
\begin{eqnarray}
\Delta m^2_{21} = 7.50 \times 10^{-5} \; \textrm{eV}^2; \quad \Delta m^2_{31} = 2.46 \times 10^{-3} \; \textrm{eV}^2;  \nonumber \\
\theta_{12} = 33.48^\circ ; \quad  \theta_{23} = 42.30^\circ ; \quad  \theta_{13} = 8.50^\circ ; \quad \delta_{CP} = 0.
\nonumber \\ \label{eq:oscparams}
\end{eqnarray}

In the analysis, we focus on the
determination of atmospheric parameters $\theta_{23}$ and $\Delta
m^2_{31}$. For simplicity, all oscillation parameters have been fixed
during the analysis; \ie no marginalization has been performed when
obtaining the allowed confidence regions. However, our conclusions are
not expected to change significantly if marginalization was performed
within the currently allowed experimental regions. Matter effects are
included in our simulations, and the matter density profile has been
chosen according to the preliminary reference Earth model~\cite{ref:PREM}.

\subsection{Considered experimental setups}

Oscillation experiments using neutrino beams produced from meson decays in flight can be divided into two main categories according to their far detector locations: on-axis (such as K2K~\cite{ref:K2K} and MINOS~\cite{ref:MINOS} and also the DUNE~\cite{ref:LBNE} and LAGUNA-LBNO~\cite{ref:LBNO} proposals) and off-axis (for instance, ongoing T2K~\cite{ref:T2K} and NO$\nu$A~\cite{ref:NOvA}).

Because of the pion-decay kinematics, the flux at an off-axis site is well localized around a given neutrino energy, with the beam spread and high-energy tail heavily reduced~\cite{ref:off-axis}. Such a design allows for a significant reduction of the backgrounds coming from neutral-current events, and guarantees that the range of values of $L/E_\nu$ is well localized around the first oscillation maximum. The price to pay is a lower beam intensity compared to an on-axis configuration. In addition, as the mean neutrino energy is lower than in an on-axis experiment, the number of events in the far detector is also lower, since the cross section is an increasing function of the energy.

On the other hand, in an on-axis neutrino oscillation experiment, the spread of the beam is larger. In principle, this allows one to determine the \emph{shape} of the oscillation probability by performing measurements at different values of $L/E_\nu$. In addition, thanks to the higher beam intensity and higher neutrino energies, large event statistics is easier to collect. As a consequence of being more energetic, though, the event sample generally contains a significant fraction of pion-production events. In addition, the high-energy tail typically produces a significant neutral-current background contaminating in particular the low-energy bins.

\begin{figure}
\begin{center}
\includegraphics[width=0.8\columnwidth]{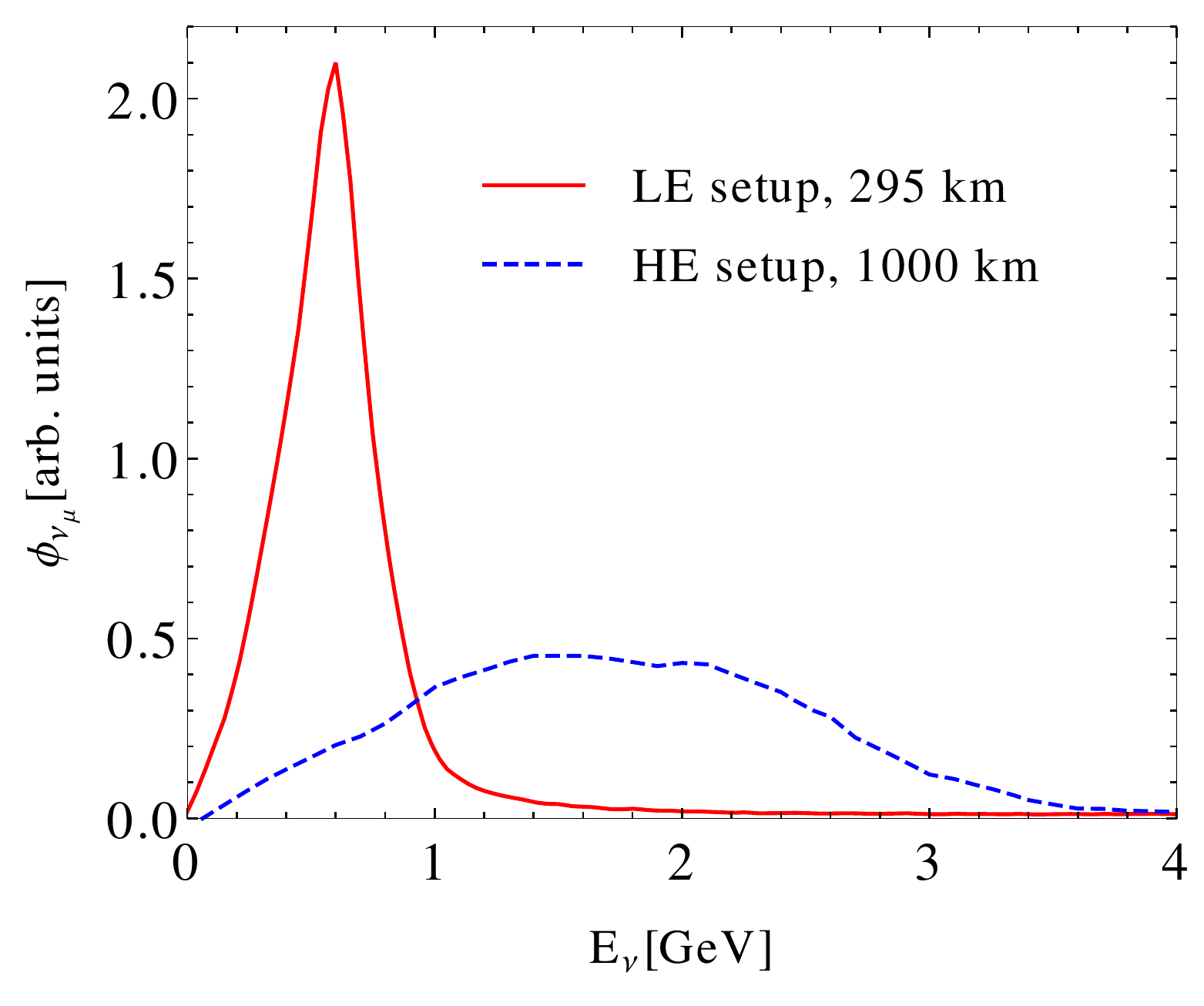}
\caption{\label{fig:fluxes}(color online). The muon neutrino fluxes (in arbitrary units) as a function of the neutrino energy for the two configurations considered in this work. The solid (dashed) line corresponds to the low-(high-)energy setup.
}
\end{center}
\end{figure}

In this article, we perform an analysis of two different neutrino-oscillation experiments: a HE setup with a broadband on-axis beam, and a LE setup with a narrow-band off-axis beam.

In the LE setup, we use the neutrino beam of Ref.~\cite{ref:Patrick2009}, being a preliminary estimate of that in the T2K experiment with the $2.5^\circ$ off-axis configuration~\cite{ref:T2K_flux}. As shown in Fig.~\ref{fig:fluxes}, it is peaked at around 600~MeV. For the distance to the far detector for the LE setup, we take 295~km, the length corresponding to the first oscillation maximum. The normalization (\ie number of protons on the target, detector size, and data-taking time) is arbitrarily set to obtain a total unoscillated CC inclusive event sample of $\mathcal{O}(5000)$ events in the energy range between 0.3 and 2 GeV, relevant to the analysis.

\begin{table}
\renewcommand{\arraystretch}{1.6}
\caption{\label{tab:events} Expected number of \emph{unoscillated} CC $\nu_\mu$ events for the low- and high-energy setups defined in the text. In addition to the total event numbers, their breakup into different reaction mechanisms, labeled as in Fig.~\ref{fig:xsec}, is also given.}
\begin{ruledtabular}
\begin{tabular}{l@{\quad} | c@{\quad} c@{\quad} c@{\quad} c@{\quad} | r@{\quad}}

        			&   QE   & $2p2h$   &   res  & DIS   & total   \\ \hline

LE (0.3--2 GeV) 	&  49\% & 28\%  &  21\%  & 2\%  & 4891    \\ \hline

HE (0.3--4 GeV) 	&  26\% & 11\% &  37\%  & 26\% & 4456    \\

\end{tabular}
\end{ruledtabular}
\end{table}

For the HE option, the neutrino flux labeled as ``650~km'' in Ref.~\cite{ref:Longhin} is used, being one of the configurations considered within the LAGUNA design study~\cite{ref:LAGUNA}. This flux has a broad peak between $\sim$1 and $\sim$2 GeV, with a non-negligible tail extending well above 3~GeV; see Fig.~\ref{fig:fluxes}. We considered several choices for the distance to the far detector in the range $L=\mathcal{O} (500\textrm{--}1000)$~km, for which the first oscillation maximum would lie within the energy range of the flux peak, and found that $L\sim 1000$~km gives optimal results in the $\nu_\mu$ disappearance channel. In the following, we discuss our results for this baseline choice only.

Since in this work we are only interested in exploring detector effects on different energy-reconstruction methods, the normalization for the HE setup is arbitrarily set to obtain a similar total number of unoscillated events as for the LE setup. However, due to the much higher neutrino energies, the composition of the inclusive event sample is very different. To illustrate it, the total number of unoscillated CC inclusive events at the far detector are given in Table~\ref{tab:events} for both setups, together with the percentages for QE, $2p2h$, res and DIS events in each sample.

In our analysis, all neutrino events with the energies between 0.2 and 8~GeV are calculated for both LE and HE setups. However, only those which are reconstructed between 0.3 and 2~GeV (4~GeV) for the LE (HE) setup are considered during the fit. The $\chi^2$ is built by binning the events in reconstructed neutrino energy, using 100~MeV bins.

\subsection{Results for the calorimetric method }
\label{sec:cal-results}

As explained in Sec.~\ref{sec:energyReconstruction}, in the calorimetric method, the neutrino energy is reconstructed by adding the energies of all observed particles in the final state, and no information on the direction of the outgoing particles is used.

In an analysis of an oscillation experiment, one has to rely on a Monte Carlo simulation to predict the ``expected'' (or ``fitted'') event rates. One of the inputs to such a simulation is detailed information on the detector capabilities. However, it is subject to uncertainties, the role of which we are going to analyze.

\begin{figure}
\begin{center}
\includegraphics[width=0.80\columnwidth]{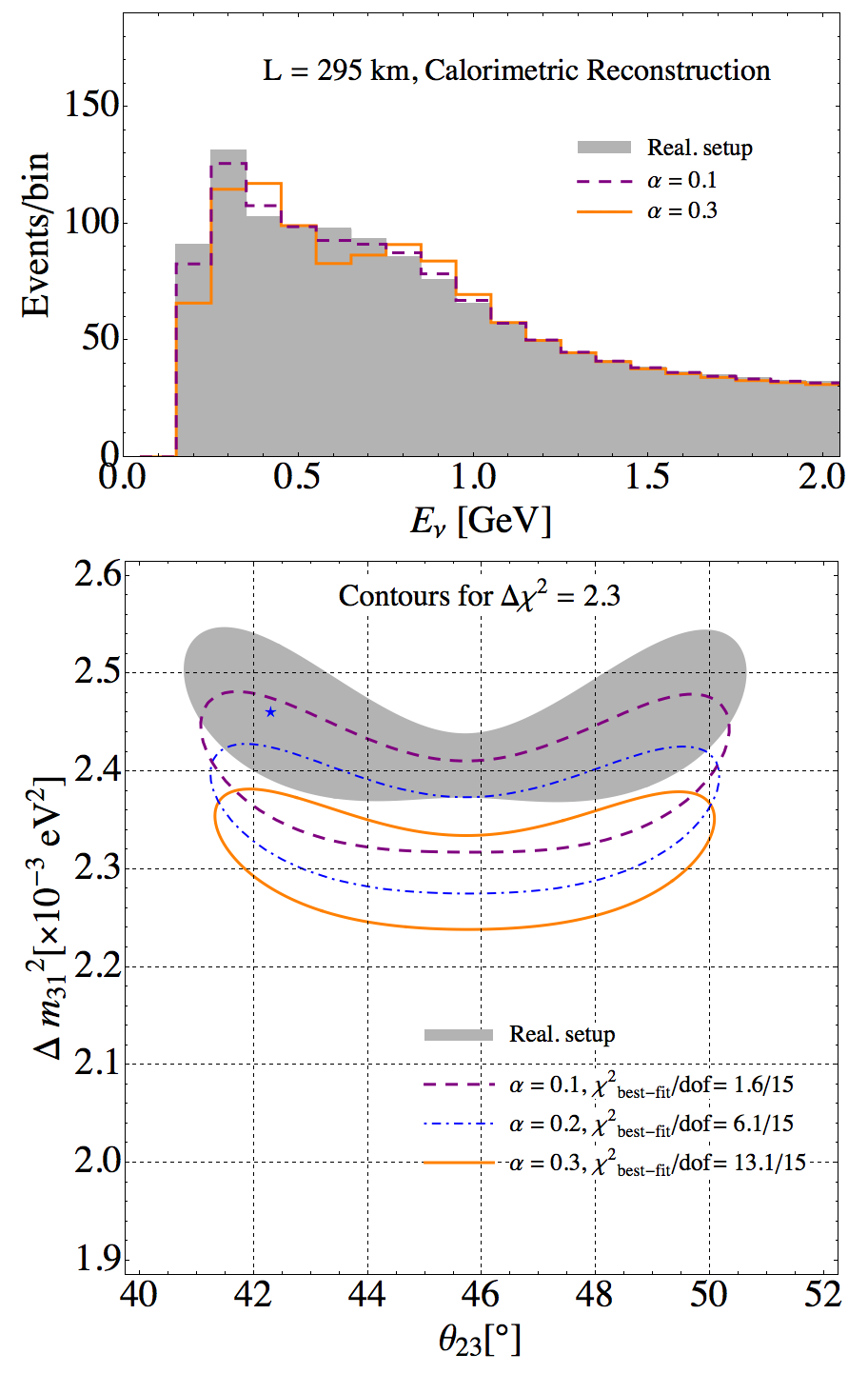}
\caption{\label{fig:T2Kcal}(color online). Results for the low-energy
  setup and the calorimetric reconstruction method.  Upper panel: Simulated charged-current event distributions
  in the far detector as a function of the reconstructed energy, for
  the oscillation parameters~\eqref{eq:oscparams}. The
  shaded histogram corresponds to a perfect estimate of detector effects (see
  Sec.~\ref{sec:scenarios}). The dashed and solid lines show the event
  rates obtained with the detector performance overestimated by 10\% and 30\%.  Lower panel: Confidence
  regions in the $(\theta_{23}, \Delta m^2_{31})$ plane, at 1$\sigma$
  C.L. (2 degrees of freedom). The shaded area corresponds to the perfect estimate of the detector effects.
  The lines show the contours obtained for the detector performance overestimated by 10\%, 20\%, and 30\% in the fit. The star indicates
  the true values of the oscillation parameters, the same for
  all confidence regions shown.  }
\end{center}
\end{figure}

\begin{figure}
\begin{center}
\includegraphics[width=0.80\columnwidth]{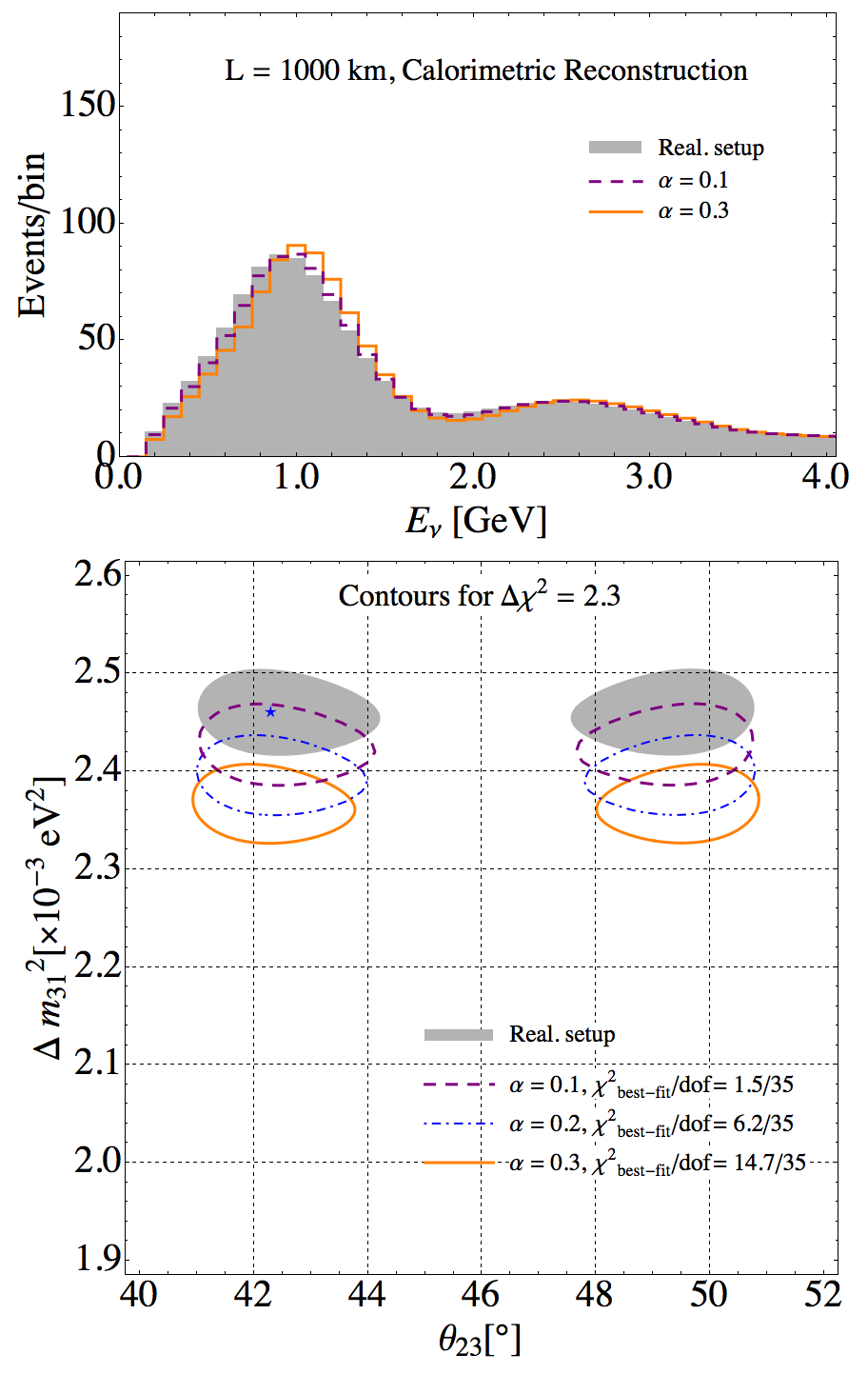}
\caption{\label{fig:LAGcal}(color online).
Same as in Fig.~\ref{fig:T2Kcal} but for the high-energy setup and the calorimetric reconstruction method.
}
\end{center}
\end{figure}

Our results for the calorimetric method are shown in Figs.~\ref{fig:T2Kcal} and~\ref{fig:LAGcal} for the LE and HE setups, respectively. In their upper panels, we present the simulated event distributions in the far detector
as a function of the reconstructed energy. The lower panels illustrate the confidence regions for fits to the atmospheric oscillation parameters. In the fits, the true event rates are computed for the oscillation parameters~\eqref{eq:oscparams} taking into account detector effects according to the realistic setup described in Sec.~\ref{sec:scenarios}.

Because it is computationally expensive to generate a given set of migration
matrices from simulated neutrino events, in our
analysis, the fitted rates are generated using a linear
combination of the matrices obtained for the realistic and perfect-reconstruction scenarios.
Allowing for a continuous transformation
of the event-rate distribution from one scenario to the other,
this method is useful to quantify when the incorrect estimation of the detector
performance starts to have a significant impact on the
fit. In practice, the fitted rates are obtained as:
\begin{equation}
N_{i}^\textrm{fit} = \sum_{X} \sum_{j} \left\{ (1-\alpha) M^{X,\textrm{ real}}_{ij}  +  \alpha M_{ij}^{X,\textrm{ perf}}  \right\} N^{X}_{j},
\label{eq:alpha}
\end{equation}
where $X$ stands for a given type of interaction (QE, $2p2h$, res, or
DIS) and $i$ and $j$ are the energy-bin indices. $M^{X, \textrm{ real}}$ and
$M^{X, \textrm{ perf}}$ denote the migration matrices obtained for the
realistic and perfect-reconstruction scenarios,
respectively. Finally, $N^X_j$ is the event rate in the energy bin $j$
for the interaction type $X$, computed without accounting for the detector effects, as in Eq.~\eqref{eq:Nev}.

In Eq.~\eqref{eq:alpha}, $\alpha$ is a \emph{purely phenomenological}
parameter, used to obtain the ``effective'' migration matrices as a
linear deformation of the two extreme scenarios under
consideration. For instance, $\alpha = 0$ means that the fitted rates
are obtained in the same way as the true rates. On the other hand,
$\alpha = 0.3$ means that the fitted rates are obtained underestimating the role of detector effects by 30\%, a substantial amount. Two examples of the event distributions obtained for different values of $\alpha$ are shown in
the upper panels in Figs.~\ref{fig:T2Kcal} and~\ref{fig:LAGcal}. Even though the distributions look very
different, all histograms have been obtained assuming the same true
values for the oscillation parameters~\eqref{eq:oscparams}.

For a given value of $\alpha$, the fitted rates are simulated for
every possible combination of $(\Delta m^2_{31}, \theta_{23})$. The
point in the $(\Delta m^2_{31}, \theta_{23})$ plane which gives a
best fit to the data is then identified, and the confidence regions
are drawn by requiring that
\begin{equation}\label{eq:confidence}
\Delta \chi^2 (\Delta m_{31}^2, \theta_{23}) \equiv \chi^2 (\Delta m^2_{31}, \theta_{23}) - \chi^2_\textrm{best-fit} < 2.30.
\end{equation}
Since the fitted rates are generated using a different set of the
migration matrices, the event distributions generally have a
different shape than the true rates, and the best fit to the data does
not necessarily coincide with the true values of oscillation parameters realized in nature. When the detector
capabilities are incorrectly estimated, the allowed confidence regions start to
drift away from the shaded areas, as shown in the lower panels in Figs.~\ref{fig:T2Kcal}
and~\ref{fig:LAGcal}. The minimum value of the $\chi^2$
obtained for each $\alpha$ is indicated in the legend,
together with the effective number of degrees of freedom (d.o.f.) in
the fit, $N_\textrm{d.o.f.}$\footnote{The effective number of degrees of freedom, being the number of bins
  used in the fit minus the number of parameters determined from the data, is typically used in tests of the
  validity of the model used to fit the data. In a real experiment,
  the minimum $\chi^2$ would receive contributions from statistical
  fluctuations of the measured quantities in each data bin, and $\chi^2_\textrm{min}/N_\textrm{d.o.f.}$ should be close to 1, if the
  model used to fit the data is correct. In our
  case, no statistical fluctuations are considered in the
  fit. Nevertheless, the use of a wrong model in data fitting
  can give a sizeable contribution to the value of
  $\chi^2_\textrm{min}/N_\textrm{d.o.f.}$. This can be estimated in our calculations
  as $\chi^2_\textrm{best-fit}/N_\textrm{d.o.f.}$. }. For $\alpha = 0$ the value of
the $\chi^2$ at the best-fit point is consistent with zero and,
therefore, is not indicated in the legend.

\subsection{Results for the kinematic method}
\label{sec:kin-results}

In the kinematic method, the neutrino energy
is reconstructed using only the direction and momentum of the produced
charged lepton in the final state, and no kinematic
information from the outgoing hadrons is needed.

\begin{figure}
\begin{center}
\includegraphics[width=0.80\columnwidth]{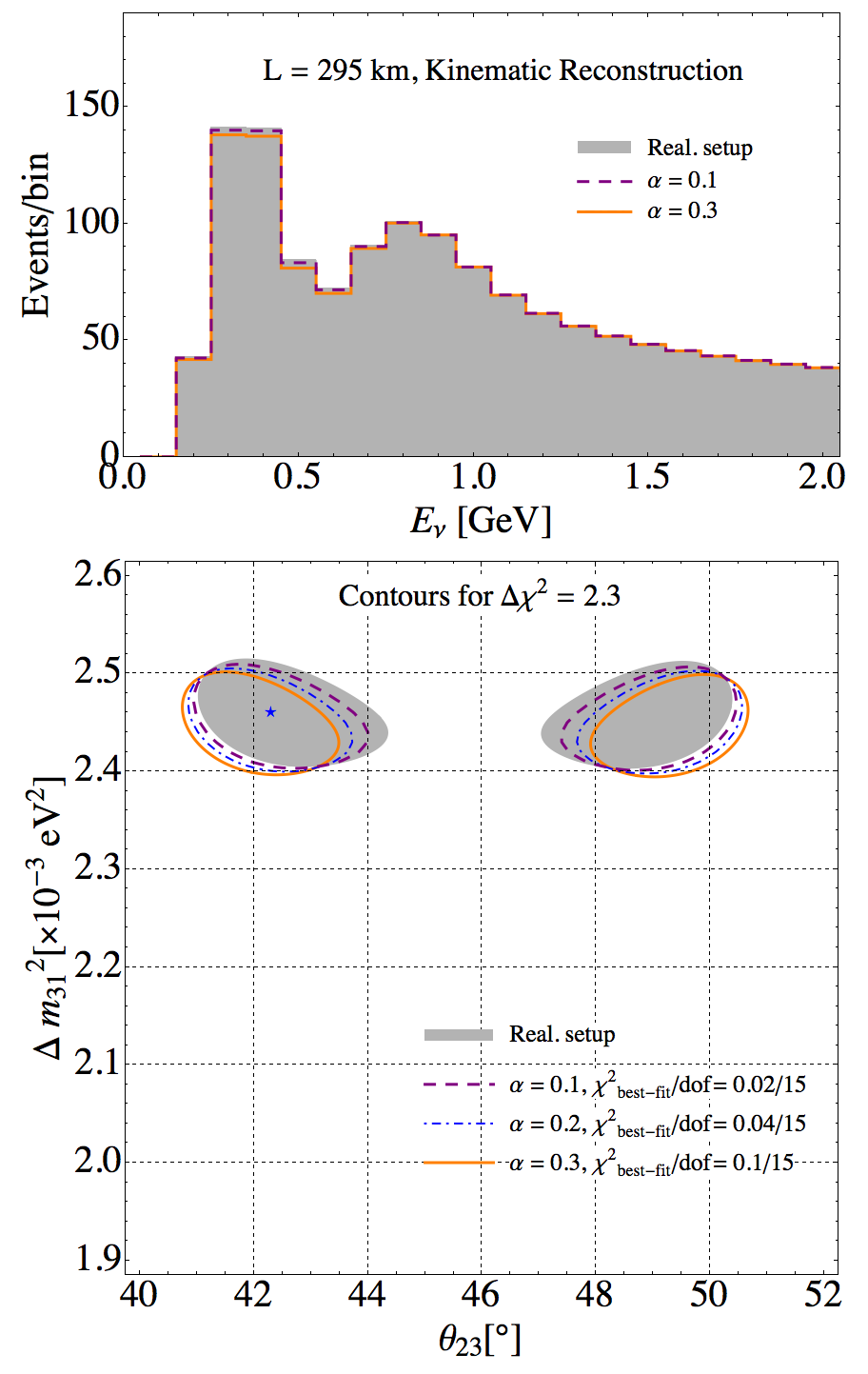}
\caption{\label{fig:T2Kkin}(color online).
Same as in Fig.~\ref{fig:T2Kcal} but using the kinematic method to reconstruct the neutrino energy.
}
\end{center}
\end{figure}

\begin{figure}
\begin{center}
\includegraphics[width=0.80\columnwidth]{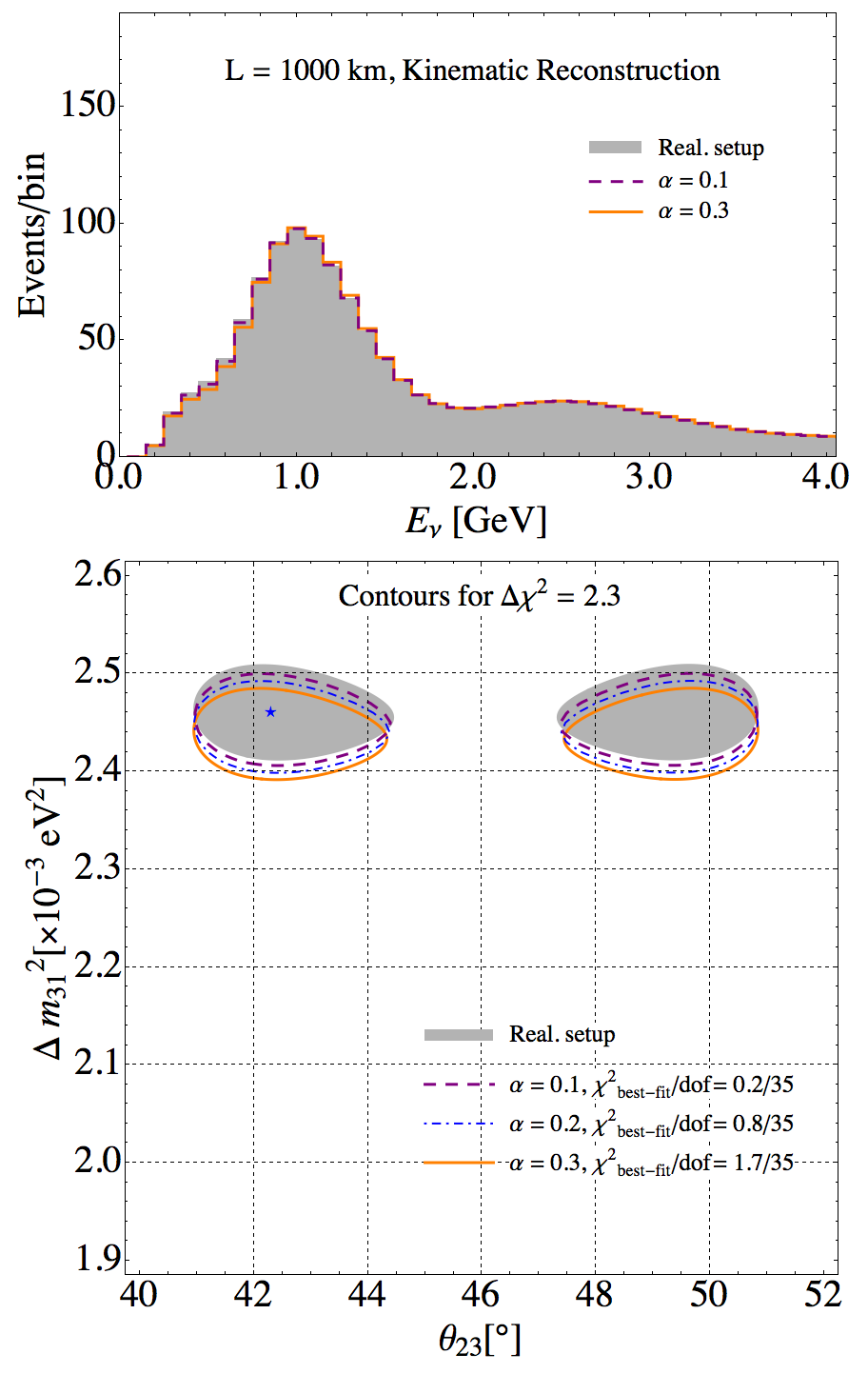}
\caption{\label{fig:LAGkin}(color online).
Same as in Fig.~\ref{fig:T2Kcal} but for the high-energy setup and the kinematic energy reconstruction.
}
\end{center}
\end{figure}

As in Sec.~\ref{sec:cal-results}, we calculate the true event distribution
 using the migration matrices accounting for realistic
 detection capabilities, while the fitted rates are obtained
following Eq.~\eqref{eq:alpha}, and the confidence regions satisfy Eq.~\eqref{eq:confidence}.

Our results for the kinematic reconstruction method are shown in Figs.~\ref{fig:T2Kkin}
and~\ref{fig:LAGkin} for the LE and HE setups, respectively. In the case of the confidence regions represented by the shaded areas, the true and fitted rates are generated using the same
set of migration matrices ($\alpha=0$), and, therefore, the best fit coincides with
the true values. When the value of $\alpha$ increases, the confidence
regions start to drift away from the shaded areas. This is a consequence of the fact that the shapes of the event
distributions are generally different from that of the true rates, even for the same oscillation probability, and the best fit for the oscillation parameters no longer coincides with the true
values.

However, it should be noted that for the kinematic reconstruction, the effect of the
mismatch between the actual and assumed detector capabilities is much
milder than for the calorimetric method. The confidence regions show a significant overlap,
even for $\alpha = 0.3$, corresponding to the detector
performance substantially overestimated.
The main reason for this is that the muon reconstruction is very precise in modern detectors,
while for electrons and hadrons this is more difficult to achieve (see also
Figs.~\ref{fig:effRes} and~\ref{fig:effMean}). It should be noted, however, that in this
study we assume that the muon track is fully contained in
the detector, regardless of its energy. Should this assumption be
relaxed, the muon momentum determination may be affected and,
as a consequence, also the neutrino-energy reconstruction.

\section{Summary}
\label{sec:summary}

It has been realized for a while now that an accurate understanding of
neutrino-nucleus scattering is an essential ingredient for
neutrino-oscillation experiments employing accelerator beams. More recently,
quantitative studies have tackled the relation between
uncertainties in cross section modeling and the resulting physics
sensitivities for oscillation measurements. For example, it was shown that for quasielastic events the energy reconstruction based
exclusively on the kinematic observables of the outgoing lepton is
susceptible to a large bias resulting from the underlying nuclear-interaction model~\cite{ref:Pilar_PRD,ref:Pilar_PRL}.
In this article, we fully rely on the nuclear model used for event generation, and have a
critical look only at the impact of realistic detector effects.

From energy conservation, it is clear that a perfect calorimeter---\ie
a detector able to measure the total energy of all reaction
products---would be free from any bias in energy reconstruction~\cite{ref:Ulrich}.
In this article, we aim at quantifying how close to the perfect scenario
the calorimetric energy reconstruction is, when finite energy resolutions,
detection efficiencies, and thresholds are accounted for.
In addition, we compare the role of realistic detection capabilities
on the calorimetric and kinematic analysis. In the latter case, we use---although in a simplistic way---the information on that the charged lepton's kinematics also for non-QE events.

We find that the kinematic reconstruction is very robust with
respect to detector effects, largely because muons are the most precisely reconstructed particles in modern neutrino detectors.
On the other hand, the actual
performance of the calorimetric analysis clearly depends on the
assumed detector performance. The employed detection capabilities
are not meant to represent any existing detector, but are indicative of the general level of performance which
can be expected. While to translate these result into a specific experiment a
detailed study of detector response---beyond the scope of this work---would be
required, many of the analysis techniques developed here should turn out to be very useful. Interestingly, we find that the kinematic reconstruction performs
well even for pion-production events, but the independence of this observation from the underlying nuclear model is to be examined.

We limit our study to the $\nu_\mu$ disappearance channel since this
can be effectively treated as a two-flavor oscillation. As a consequence, the effects
of energy resolution and misreconstruction are directly
related to the precision of the $\Delta m^2$ determination and a shift of the value
of $\Delta m^2$, respectively.  We use a phenomenological
parametrization to interpolate between a perfect and realistic
detector. This allows us to conclude that, overall, the detector
response---in terms of efficiencies, resolutions, and thresholds for
individual particles---has to be understood at a 10\% level or better,
to avoid a significant bias in the measurement of $\Delta m^2$. For
the kinematic analysis, these requirements are much less stringent, but
uncertainties of the nuclear model---not considered here---present a challenge,
as shown previously in the literature.

In summary, while the calorimetric reconstruction may be less sensitive
to the underlying nuclear model, it is strongly affected by
detector effects, typically leading to energy underestimation.
On the other hand, the kinematic method of neutrino energy reconstruction is
much less challenging for the detector design, but it strongly relies on an
accurate understanding of neutrino-nucleus interactions.
One needs to keep in mind that the quantitative details of our conclusions
may be specific to this work, owing to the underlying detector assumptions.
However, their qualitative aspects can be expected to hold for a variety of experiments.

As a final remark, we observe that the results presented in this article are subject to nuclear-model uncertainties, likely to be largest for $2p2h$ processes. Their estimate is left to be quantified in our future studies.

\begin{acknowledgments}
The work of A.M.A., C.M.J., and C.M. was supported by the National Science Foundation under Grant No. PHY-1352106.
Fermilab is operated by the Fermi Research Alliance under Contract No. \protect{DE-AC02-07CH11359} with the
U.S. Department of Energy. P.C. acknowledges partial support from the European Union FP7 ITN INVISIBLES
(Marie Curie Actions, Grant No. PITN- GA-2011- 289442). P.H. is supported by the U.S. Department of Energy under Contract No. \protect{DE-SC0013632}. The work of D.M. and E.V. was supported by MIUR (Italy) under the program
``Futuro in Ricerca 2010 (RBFR10O36O)''. E.V. acknowledges the hospitality and
support from Center for Neutrino Physics of Virginia Tech.
\end{acknowledgments}

\end{document}